\documentclass[11pt]{article}
\usepackage[cp850]{inputenc}
\usepackage{colortbl}
\usepackage{epsf,psfig}
\usepackage{amsfonts,amssymb,latexsym}
\newcounter{fig}

\newcounter{gif}

\newcommand{\beq}{\begin{eqnarray*}}
\newcommand{\eeq}{\end{eqnarray*}}
\newcommand{\beqn}{\begin{eqnarray}}
\newcommand{\eeqn}{\end{eqnarray}}
\newcommand{\mref}[1]{$(\ref{#1})$}  % reference for mathematical formul\ae
\renewcommand{\cellcolor}[3]{\multicolumn{1}{>{\columncolor[#1]{#2}}c}{#3}}
\newcommand{\N}{\mathbb{N}}

\def\sqr#1#2{{\vcenter{\vbox{\hrule height.#2pt          % square
            \hbox{\vrule width.#2pt height#1pt \kern#1pt
                  \vrule width.#2pt}\hrule height.#2pt}}}}

\newcommand{\cet}[1]{\mid \! #1 \, \rangle}

\newcommand{\W}{$\cal W$}
\newcommand{\lb}{&& \!\!\!\!\!\!\!\!\!\!\! \textstyle}
\begin{document}

\title{Notes on Generalised Nullvectors in logarithmic CFT}
\author{Holger Eberle\footnote{email: {\tt eberle@th.physik.uni-bonn.de}} \hspace{0.1cm}
and \hspace{-0.1cm} Michael Flohr\footnote{email: {\tt flohr@th.physik.uni-bonn.de}}\vspace{0.3cm} \\
Physikalisches Institut der Universit\"at Bonn\\ 
Nu\ss allee 12, 53115 Bonn, Germany}

\date{}

\maketitle

\vspace{-8cm}
\begin{flushright}
BONN-TH-2005-07\\
hep-th/0512254
\end{flushright}
\vspace{+7cm}

%\tableofcontents
%\addcontentsline{toc}{section}{\protect\numberline{4}{...}           % Hinzuf\"ugen ins Inhaltsverzeichnis

%%%%%%%%%%%%%%%%%%%%%%%%%%%%%%%%%%%%%%%%%%%%%

\begin{abstract}
In these notes we discuss the procedure
how to calculate nullvectors in general indecomposable
representations which are encountered in logarithmic
conformal field theories.
In particular, we do not make use of any of the 
restrictions which have been imposed in
logarithmic nullvector calculations up to now,
especially the quasi-primarity of all Jordan cell fields.

For the quite well-studied
$c_{p,1}$ models
we calculate examples of logarithmic null\-vectors 
which have not been accessible
to the older methods and recover the known
representation structure.
Furthermore, we calculate logarithmic
nullvectors in the up to now
almost unexplored general augmented $c_{p,q}$ models
and use these to find bounds on their
possible representation structures.
\end{abstract}

%%%%%%%%%%%%%%%%%%%%%%%%%%%%%%%%%%%%%%%%%%%%%

\section{Introduction}

In recent years the study of conformal field theories (CFTs) which
also exhibit indecomposable structures in part of their
representations has become an interesting and promising
topic of research. A variety of
applications of this sensible generalisation of ordinary CFTs,
which are also commonly known as logarithmic CFTs,
have already surfaced in statistical physics (e.g.\ \cite{Sal92,GFN97,Rue02,PRG01}),
in Seiberg Witten theory (e.g.\ \cite{Flohr04}) and even in string theory
(e.g.\ \cite{LLM03,Bakas02}).
For a more complete survey of applications pursued so far as well as an
introduction to the field see 
\cite{Flohr01,Gab01,Kawai02,FG05}.

Up to now the main focus of research has been put on a special class
of logarithmic CFTs, the $c_{p,1}$ models. The representation theory
of their rank $2$ indecomposable representations has been analysed
completely in \cite{Roh96,GK96,GK96b,GK98} 
and a  thorough under\-standing
of the representations of the modular group corresponding
to the enlarged triplet \W-algebra \cite{Kau90} of these models  
has been reached in \cite{Flohr95,Flohr96,FG05}.
Especially the $c_{2,1} = -2$ model has been understood very well
as it is isomorphic to a free field construction of the
symplectic fermions \cite{Kau00,KF02}.

But going beyond representation theory we find that
the calculation of explicit
correlation functions proves to be much more intricate and
tedious than in the ordinary CFT case \cite{Flohr01b,FK05}. The construction
of nullvectors, the
key tool in CFT for the calculation of correlation functions,
has already been addressed in \cite{Flohr97,Flohr00,Mog00,Mog02} for the case
of indecomposable representations.
However, the type of these logarithmic nullvectors calculated so far
only describes a very special case. Already in the
$c_{p,1}$ models the generic logarithmic nullvectors are
beyond the scope of this procedure.

On the other hand, the $c_{p,1}$ models are only quite special
representatives of the general class of augmented $c_{p,q}$ models
\cite{Flohr96}.
Although these models have already been addressed in 
some papers (see e.g.\ \cite{FM05}),
not much is known yet, neither about higher rank representations
nor about nullvectors nor about correlation functions. There
are, however, good indications that exactly these models
might describe important statistical systems, such as perlocation.
Especially the augmented $c_{2,3} = 0$ model seems to be of high interest
in this respect \cite{PRG01}.

The main goal of this paper is to show how to calculate logarithmic 
nullvectors in general, to give explicit examples and to use the information
about the existence of nontrivial logarithmic nullvectors
to explore the unknown structure of a more general class
of logarithmic CFTs, namely the augmented $c_{p,q}$ models.

In section \ref{section1} we give a short review of the
special version of nullvector calculations which has been performed
up to now in logarithmic models. In section \ref{section2}
we discuss the limitations of this ansatz and propose a
more general procedure which is cap\-able of calculating
all lowest logarithmic nullvectors in the $c_{p,1}$ models.
In particular, our new method does not rely on the quasi-primarity of all
Jordan cell fields.
In section \ref{section3} we then briefly introduce
what we mean by generic augmented $c_{p,q}$ models.
We use the methods of the preceding sections with slight
modifications in order to obtain constraints on what
embedding structures possibly yield rank $2$ indecomposable
representations in these models. For this we will concentrate
on the ``smallest'' model exhibiting this more generic
behaviour, the augmented $c_{2,3}=0$ model. But any emerging
structures should immediately generalise to any $c_{p,q}$ model.

%%%%%%%%%

\section{Jordan cells on lowest weight level\label{section1}}

Let us briefly recall the construction of nullvectors
in a logarithmic representation in which the states 
of the Jordan cell are all lowest weight states \cite{Flohr97,Flohr00,Flohr01}. 
We will especially clarify the respective procedure in \cite{Flohr01} and 
give a proof of the proposed logarithmic nullvector conditions.
In the following, we will concentrate on Virasoro representations 
and try to keep close to the notations of \cite{Flohr01}.

A Jordan cell of lowest weight states with weight $h$ of rank $r$
is spanned by a basis of states
\beq
| h; n \rangle &=& \frac{1}{n!} \; \theta^n \; | h \rangle  \qquad \qquad \forall \; n=0,\dots, r-1
\eeq
on which the action of the Virasoro modes is given by
\beq
L_0 |h;n\rangle &=& h \, |h;n\rangle + (1- \delta_{n,0}) \, |h;n-1\rangle \nonumber \\
& \equiv & (h+\partial_{\theta} ) |h;n\rangle \; , \nonumber \\
L_p |h;n\rangle &=& 0 \qquad \qquad \forall \; p>0 \; .
\eeq
As already defined in \cite{Flohr01}, $\theta$ is a nilpotent variable with $\theta^r=0$
and a handy tool to organize the Jordan cell states with the same weight. 
Due to their almost primary behaviour with the only defect of an 
additional term in the indecomposable $L_0$ action we will call the
$|h;n\rangle$ logarithmic primary. We also note that $|h;0\rangle$ is indeed a true primary state.

Using this information about the action of the Virasoro modes 
one can easily deduce that the action of any function
of the Virasoro zero-mode and the central charge operator $f(L_0,C)$ on
such a Jordan cell state is given by \cite{Flohr01}
\beqn \label{eqn1}
f(L_0,C) \; |h;n \rangle &=& \sum_{k=0}^n \, \frac{1}{k!} \left( \frac{\partial^k}{\partial h^k} f(h,c) \right) \, |h;n-k \rangle \; .
\eeqn

On the other hand, calculation of logarithmic two-point-functions
yields the following Shapovalov form of these Jordan cell states \cite{Flohr01b}
\beqn \label{shap}
\langle h;k | h;l \rangle &=&
\left\{
\begin{array}{cc}
0 & \forall \; l+k < r-1 \\
1 & \forall \; l+k = r-1 \\
D_{l+k-r+1} & \forall \; l+k > r-1
\end{array}
\right. 
\eeqn
for constant $D_j$, $j=1,\dots,r-1$.

We now want to construct vectors which are null on the whole logarithmic
representation. Following \cite{Flohr01} we choose the general ansatz
\beqn \label{ansatz1}
|\chi_{h,c}^{(n)} \rangle &=& \sum_{j=0}^{r-1} \sum_{|{\bf n}|=n} 
\; b_j^{\bf n} (h,c) \, L_{-{\bf n}} \, |h;\partial_{\theta}^j a(\theta) \rangle
\eeqn
with $a(\theta) = \sum_{i=0}^{r-1} \, a_i \, \frac{1}{i!} \, \theta^i$ and the 
usual multi-index notation for the modes $L_{{\bf n}}$. Choosing the
states $|h;j \rangle$ instead of $|h;\partial_{\theta}^j a(\theta) \rangle$ on the
right hand side would have yielded the same generality of the ansatz and in the
end the same set of solutions. We prefer this special ansatz, however, for two reasons.
First of all it already incorporates
our knowledge that in this form of logarithmic representations there will only be
a nontrivial action of the $L_0$ modes in the end of the nullvector calculation
and that this is given by derivatives wrt $\theta$ onto lower ranks.
This justifies the $\partial_{\theta}^j$ part of the ansatz.
On the other hand, by including lower orders of $\theta^i$ into $a(\theta)$
we solve for nullvectors of lower rank subrepresentations at the same time.
For this we will always treat the $a_m$ as arbitrary parameters.

Now, we can calculate the  level $n$ nullvector conditions for arbitrary 
$k$ and ${\bf n}_l$, $|{\bf n}_l|=n$, as follows (the index $l$
indicates a suitable enumeration of the multi-indices ${\bf n}$):
\beqn \label{null_cond}
\lefteqn{\langle h; r-1-k | L_{{\bf n}_l} \,  |\chi_{h,c}^{(n)} \rangle} \nonumber \\
&=& \sum_{j=0}^{r-1} \sum_{|{\bf n}|=n} \, b_j^{\bf n} (h,c) \,
\sum_{m=j}^{r-1} \, a_m \langle h; r-1-k | L_{{\bf n}_l} \,  L_{-{\bf n}} \, |h;m-j \rangle \nonumber \\
&=& \sum_{j=0}^{r-1} \sum_{|{\bf n}|=n} \sum_{m=j}^{r-1}  
\, a_m \, b_j^{\bf n} (h,c) \, \langle h; r-1-k | \sum_{t=0}^{m-j} \,\frac{1}{t!} \left( \frac{\partial^t}{\partial h^t} 
f_{{\bf n}_l,{\bf n}} (h,c) \right) \, |h;m-j-t \rangle \nonumber \\
&=& \sum_{m=0}^{r-1} \, a_m \, \sum_{j=0}^{m} \sum_{|{\bf n}|=n}  \sum_{t=0}^{m-j}
\, b_j^{\bf n} (h,c) \,\frac{1}{t!} \left( \frac{\partial^t}{\partial h^t} f_{{\bf n}_l,{\bf n}} (h,c) \right) \nonumber \\
&& \qquad \qquad \qquad \qquad \qquad \quad \times \,
\left(\delta_{t,m-j-k} +  \sum_{s=0}^{m-j-k-1} \delta_{s,t} \, D_{m-j-k-s}  \right) \nonumber \\
&=:& \sum_{m=0}^{r-1} \; a_m \, {\cal A}^{\prime}_{{\bf n}_l} (m,k) \; ,
\eeqn
where in the second step we used \mref{eqn1} due to the logarithmic primarity
of the Jordan cell ground states $|h;m \rangle$
as well as in the third step the Shapovalov form \mref{shap}.
As we want to keep the $a_m$ as arbitrary parameters the nullvector conditions on $|\chi_{h,c}^{(n)} \rangle$
are equivalent to the identical vanishing of all ${\cal A}^{\prime}_{{\bf n}_l} (m,k)$.

Let us calculate the terms proportional to $\delta_{t,m-j-k}$ in ${\cal A}^{\prime}_{{\bf n}_l} (m,k)$ first:
\beq
\lefteqn{\sum_{j=0}^{m} \sum_{|{\bf n}|=n}  \sum_{t=0}^{m-j}
\, b_j^{\bf n} (h,c) \,\frac{1}{t!} \left( \frac{\partial^t}{\partial h^t} 
f_{{\bf n}_l,{\bf n}} (h,c) \right) \; \delta_{t,m-j-k}} \nonumber \\
&=& \sum_{j=0}^{m-k } \sum_{|{\bf n}|=n} \, b_j^{\bf n} (h,c) \,\frac{1}{(m-k-j)!} 
\left( \frac{\partial^{m-k-j}}{\partial h^{m-k-j}} f_{{\bf n}_l,{\bf n}} (h,c) \right) \nonumber \\
&=:& {\cal A}_{{\bf n}_l} (m-k) \; .
\eeq
It is important to notice that the ${\cal A}_{{\bf n}_l} (m-k)$ indeed only depend on the difference $m-k$.

We now show that the vanishing of the ${\cal A}_{{\bf n}_l} (m-k)$ is necessary and already sufficient
for the vanishing of the nullvector conditions in \mref{null_cond} by using complete
induction over $m-k$. For $m-k=0$ we trivially find ${\cal A}^{\prime}_{{\bf n}_l} (m,k) = {\cal A}_{{\bf n}_l} (m-k)$.
This can easily be inferred from \mref{null_cond} noting that in the fourth line the summation over $t$
actually only runs from $0$ to $m-j-k$, consequently the one over $j$ only from $0$ to $m-k$.
On the other hand we find for general $m-k$
\beq
\lefteqn{\!\!\!\!\!\!\!\!{\cal A}^{\prime}_{{\bf n}_l} (m,k) - {\cal A}_{{\bf n}_l} (m-k)} \nonumber \\
&=&  \sum_{j=0}^{m} \sum_{|{\bf n}|=n}  \sum_{t=0}^{m-j-k-1} \, b_j^{\bf n} (h,c) 
\,\frac{1}{t!} \left( \frac{\partial^t}{\partial h^t} f_{{\bf n}_l,{\bf n}} (h,c) \right) \, D_{m-j-k-t} \nonumber \\
&=& \sum_{j=0}^{m-k-1} \sum_{|{\bf n}|=n}  \sum_{t=j}^{m-k-1} \, b_j^{\bf n} (h,c) 
\,\frac{1}{(t-j)!} \left( \frac{\partial^{t-j}}{\partial h^{t-j}} f_{{\bf n}_l,{\bf n}} (h,c) \right) \, D_{m-k-t} \nonumber \\
&=& \sum_{t=0}^{m-k-1} \, D_{m-k-t} \, \sum_{|{\bf n}|=n}  \sum_{j=0}^{t} \, b_j^{\bf n} (h,c) 
\,\frac{1}{(t-j)!} \left( \frac{\partial^{t-j}}{\partial h^{t-j}} f_{{\bf n}_l,{\bf n}} (h,c) \right) \nonumber \\
&=&  \sum_{t=0}^{m-k-1} \, D_{m-k-t} \, {\cal A}_{{\bf n}_l} (t) \; .
\eeq
But this vanishes due to the induction assumption ${\cal A}_{{\bf n}_l} (t) = 0$ for all $t < m-k$.
Hence, the vanishing of ${\cal A}^{\prime}_{{\bf n}_l} (m,k)$ is equivalent to the vanishing of ${\cal A}_{{\bf n}_l} (m-k)$
and therefore the statement is proven.

Now, if we regard these calculations with $a_{r-1}$ as the only non-vanishing parameter
we see that we still retain the whole set of conditions, ${\cal A}_{{\bf n}_l} (r-1-k) = 0$ for all $k=0, \dots, r-1$.
On the other hand, taking another $a_i$ with $i<r-1$ as the only non-vanishing parameter
automatically yields the respective set of conditions for a rank $i+1$ nullvector,
a nullvector of a rank $i+1$ logarithmic subrepresentation.
This indeed justifies our chosen ansatz as well as keeping the parameters $a_m$ arbitrary.

Furthermore, this calculation shows that any nullvector wrt a (logarithmic or irreducible) representation
is automatically a nullvector to any larger logarithmic representation
containing the former one as a subrepresentation.

The third fact we would like to stress is that, generically, the nullvector of some rank $r$
logarithmic representation is not a pure descendant of the Jordan cell state with rank
index $r$, but always contains descendants of the
other Jordan cell states with lower rank index, including the groundstate of the irreducible representation.

%%%%%%%%%

\section{Logarithmic nullvectors for $\mathbf{c_{p,1}}$\label{section2}}

Already for the well-studied $c_{p,1}$ models, however,
the representations with Jordan cells on the lowest
level analysed in the preceding section are not the
end of the story but rather only very special cases \cite{GK96}.
For the generic rank $2$ logarithmic representations
in these models one needs a generalised way of calculating
logarithmic nullvectors, which we will develop in the following.
We will introduce these representations using the notational conventions of
\cite{GK96} and will then transfer to notations which
are more adapted to our procedure.

\begin{figure}
\centering \leavevmode
\psfig{file=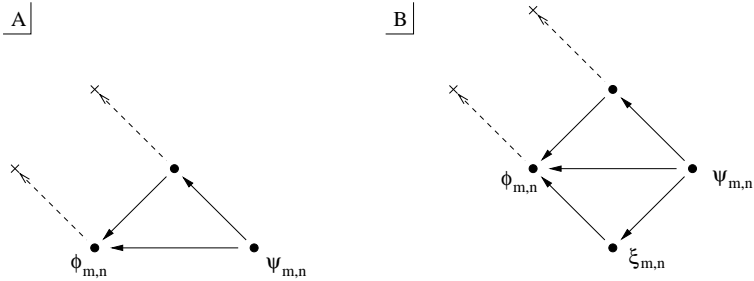,width=10cm}
\caption{Types of rank $2$ representations in $c_{p,1}$ models}
\label{rep_on_border}
\end{figure}

In figure \ref{rep_on_border} we have depicted the two possible
types of rank $2$ Virasoro representations which appear for $c_{p,1}$ 
as calculated in \cite{GK96}. The dots
correspond either to generating fields, i.e.\ fields
which are not describable as descendants of some other field, or
to singular descendants of these which, although they are
null in the module of their parent field, have a non-vanishing
Shapovalov form with some other field of the whole
rank $2$ representation. The crosses, on the other hand,
represent true nullvectors of the whole rank $2$ representation.
And finally, the arrows indicate which fields may be reached
by the application of some polynomial in the Virasoro modes.
But as remarked in the last section, nullvectors which are built in part
on the second (or any higher) Jordan cell field always have to
contain contributions from descendants of the primary field 
(and possibly lower Jordan cell fields) as well.
The corresponding arrows have to be understood in this way.
The naming of the fields follows the convention in \cite{GK96},
where the indices $m$, $n$ refer to the Kac labels corresponding
to the weight of the Jordan cell fields.

The case of a Jordan cell built solely
on logarithmic primary fields which we discussed in section \ref{section1} and 
which corresponds to case {\bf A}
in figure \ref{rep_on_border}
is just the exceptional case for the lowest lying 
representations (``lowest'' in the sense of integer differences between the weights
of the cyclic states).
The generic rank $2$ representation which is shown as case {\bf B} in figure \ref{rep_on_border}
contains an extra field $\xi_{m,n}$ with lower weight
than the Jordan cell which serves as a parent to the primary field $\phi_{m,n}$,
the generator of the
irreducible subrepresentation of the Jordan cell. 
Certainly, $\phi_{m,n}$ needs to be a singular descendant of $\xi_{m,n}$
in order to be primary. It is normalised such that the coefficient
of the $L_{-1}^l$ term is $1$.
Furthermore, the second
field building the Jordan cell, the so-called ``log-partner'' $\psi_{m,n}$, 
is not logarithmic primary any more
but is mapped to $\xi_{m,n}$ by some
polynomial of positive Virasoro modes. As argued in \cite{GK96}, if there
is no additional null\-vector on $\xi_{m,n}$ on a level lower than $\phi_{m,n}$,
this polynomial can be chosen as monomial such that
\beq
(L_1)^l \psi_{m,n} =\beta \: \xi_{m,n} &\qquad & L_p \psi_{m,n} = 0 \quad \forall \; p \ge 2
\eeq
for $l=h(\phi_{m,n})-h(\xi_{m,n})$ and constant $\beta$ depending on the representation.
In this setting $L_1$ maps $\psi_{m,n}$ to the unique 
$l-1$ descendant $\xi^D$ of $\xi_{m,n}$ with
\beqn \label{unique}
(L_1)^{l-1} \xi^D =\beta \: \xi_{m,n} &\qquad & L_p \xi^D = 0 \quad \forall \; p \ge 2 \; .
\eeqn

This kind of representation requires a more general ansatz of logarithmic
nullvectors. Loosing the prerequisite of logarithmic primarity of all
Jordan cell fields we cannot assume that only polynomials in the Virasoro
null-mode and the central charge operator contribute to the
matrix elements in the calculation of the nullvector conditions ---
we now have to take into account operators $(L_{-1} L_1)^j$, $j>0$, as well.
Hence, the relation between the nullvector polynomials on the different
Jordan cell states is not governed by the action of $L_0$ and, thus,
derivatives by $\theta$ alone. An ansatz of the form \mref{ansatz1} is not 
reasonable any more.

Instead, we propose the more general ansatz for the generic rank $2$ representation
\beqn \label{ansatz2}
|\chi_{h,c}^{(n)} \rangle &=& \sum_{|{\bf n}|=n} \; b_1^{\bf n} (h,c) \, L_{-{\bf n}} \, |h;1 \rangle + 
\sum_{|{\bf m}|=n+l} \; b_0^{\bf m} (h,c) \, L_{-{\bf m}} \, |h-l \rangle \; .
\eeqn
Here we choose a notation close to section \ref{section1} describing a state by its
weight and enumerating Jordan cell states according to the $L_0$ action
\beq
L_0 |h;n\rangle &=& h \, |h;n\rangle + (1- \delta_{n,0}) \, |h;n-1\rangle \; .
\eeq
The ansatz \mref{ansatz2} certainly incorporates general level $n$ descendants
of $\phi_{m,n} = |h;0 \rangle$ as $|h;0 \rangle$ is just a level $l$ descendant of 
$\xi_{m,n} = |h-l \rangle$
itself. However, we need this more general ansatz \mref{ansatz2} because
building descendants only on the Jordan cell states we would miss out several states
of the rank $2$ representation which are descendants of level $n+l$ of 
$\xi_{m,n} = |h-l \rangle$, but cannot be written as descendants of
$\phi_{m,n} = |h;0 \rangle$.

This ansatz leads to the following complete set of nullvector conditions
\beqn \label{new_cond1}
0 & \stackrel{!}{=} & \langle h;1 |  L_{{\bf n}_i} |\chi_{h,c}^{(n)} \rangle \nonumber \\
&=&  \sum_{|{\bf n}|=n} \; b_1^{\bf n} (h,c) \, \langle h;1 |  L_{{\bf n}_i} \, L_{-{\bf n}} \, |h;1 \rangle + 
\sum_{|{\bf m}|=n+l} \; b_0^{\bf m} (h,c) \, \langle h;1 |  L_{{\bf n}_i} \, L_{-{\bf m}} \, |h-l \rangle \nonumber \\
&=& \sum_{|{\bf n}|=n} \; b_1^{\bf n} (h,c) \, \langle h;1 |  
F^{(1)}_{{\bf n}_i,{\bf n}} (L_{-1}^l\, L_1^l, L_0, C) \, |h;1 \rangle  \nonumber \\
&& \qquad \qquad + \sum_{|{\bf m}|=n+l} \; b_0^{\bf m} (h,c) \, \langle h;1| 
F^{(2)}_{{\bf n}_i,{\bf n}} (L_{-1}^l, L_{-1}^l \, L_0^s, L_{-1}^l \, C^t) \, |h-l \rangle
\eeqn
as well as 
\beqn \label{new_cond2}
0 & \stackrel{!}{=} & \langle h-l |  L_{{\bf m}_j} |\chi_{h,c}^{(n)} \rangle \nonumber \\
&=&  \sum_{|{\bf n}|=n} \; b_1^{\bf n} (h,c) \, \langle h-l |  L_{{\bf m}_j} \, L_{-{\bf n}} \, |h;1 \rangle + 
\sum_{|{\bf m}|=n+l} \; b_0^{\bf m} (h,c) \, \langle h-l |  L_{{\bf m}_j} \, L_{-{\bf m}} \, |h-l \rangle \nonumber \\
&=& \sum_{|{\bf n}|=n} \; b_1^{\bf n} (h,c) \, \langle h-l |  
F^{(3)}_{{\bf m}_j,{\bf n}}(L_1^l, L_0^s \, L_1^l, C^t \, L_1^l) \, |h;1 \rangle    \nonumber \\
&& \qquad \qquad \qquad \qquad + \sum_{|{\bf m}|=n+l} \; b_0^{\bf m} (h,c) \, \langle h-l | 
F^{(4)}_{{\bf m}_j,{\bf n}} (L_0, C) \, |h-l \rangle \; ,
\eeqn
for any $s,t>0$.
Several remarks are necessary. The functions $F^{(1)}, \dots , F^{(4)}$ indicate what polynomials of
Virasoro generators we can reduce the interior of the above matrix elements to by successively using 
the Virasoro algebra, the level matching condition 
as well as properties of the states which these modes are applied to.
Although we are not able to reduce these to polynomials solely of $L_0$ and $C$ as in section \ref{section1},
these properties make possible a fair amount of reduction to functions which are polynomials
only of specific combinations of the four operators $L_{-1}$, $L_0$, $L_1$ and $C$.
More specifically the function $F^{(2)}$ actually only depends on terms proportional to 
$L_{-1}^l$, $L_{-1}^l \, L_0$, $L_{-1}^l \, L_0^2$, \dots as well as 
$L_{-1}^l \, C$, $L_{-1}^l \, C^2$, \dots This follows from the fact
that to the right this function acts on a primary field, to the left however on a field
which vanishes under the action of $L_p$, $p \ge 2$, 
and whose weight is just at level $l$ above $|h-l \rangle$.

As discussed earlier, we do not retain such nice interrelations between the nullvector
polynomials as in section \ref{section1} which could be cast into the $\theta$ calculus.
But we can still find remnants of such relations as e.g.\ by
looking at the nullvector conditions given by the application of level $n$
descendants of $|h;0 \rangle$ onto the nullvector ansatz
\beqn \label{new_cond3}
0 & \stackrel{!}{=} & \langle h;0 |  L_{{\bf n}_j} |\chi_{h,c}^{(n)} \rangle \nonumber \\
&=&  \sum_{|{\bf n}|=n} \; b_1^{\bf n} (h,c) \, \langle h;0 |  L_{{\bf n}_j} \, L_{-{\bf n}} \, |h;1 \rangle + 
\sum_{|{\bf m}|=n+l} \; b_0^{\bf m} (h,c) \, \langle h;0 |  L_{{\bf n}_j} \, L_{-{\bf m}} \, |h-l \rangle \nonumber \\
&=& \sum_{|{\bf n}|=n} \; b_1^{\bf n} (h,c) \, \langle h;0 |  
F^{(1)}_{{\bf n}_j,{\bf n}} (L_{-1}^l\, L_1^l, L_0, C) \, |h;1 \rangle  \nonumber \\
&& \qquad \qquad + \sum_{|{\bf m}|=n+l} \; b_0^{\bf m} (h,c) \, \langle h;0| 
F^{(2)}_{{\bf n}_j,{\bf n}} (L_{-1}^l, L_{-1}^l \, L_0^s, L_{-1}^l \, C^t) \, |h-l \rangle \nonumber \\
&=& \sum_{|{\bf n}|=n} \; b_1^{\bf n} (h,c) \, \langle h;0 |  
F^{(1)}_{{\bf n}_j,{\bf n}} (L_{-1}^l\, L_1^l, L_0, C) \, |h;1 \rangle \; .
\eeqn
These conditions are clearly a subset of the conditions \mref{new_cond2} as $|h;0 \rangle$
is just a descendant of $|h-l \rangle$. Now we make use of the Shapovalov form \mref{shap}
to deduce that the only terms contributing to the matrix elements in \mref{new_cond3}
can come from contributions of $F^{(1)}_{{\bf n}_j,{\bf n}} (L_{-1}^l\, L_1^l, L_0, C) \, |h;1 \rangle$
which are proportional to $|h;1 \rangle$.
But then we can insert these vanishing equations back into \mref{new_cond1} concluding
that the terms in \mref{new_cond1} proportional to $D_1$ (of the Shapovalov form)
already vanish on their own --- a consequence of a subset of the relations \mref{new_cond2}.
This is a reminiscence of the fact that in section \ref{section1} the conditions 
${\cal A}^{\prime}_{{\bf n}_l} (m,k)$ can be split into the conditions
${\cal A}_{{\bf n}_l} (m-k)$ which only depend on the difference $m-k$.
Hence we can conclude that any logarithmic nullvector of the proposed kind does not depend
on the constants of the Shapovalov form.

Now one can put the Virasoro-algebraic calculations on the computer
and solve the resulting equations for
possible logarithmic nullvectors. 
Details about the implementation can be found in appendix \ref{computer}.
We have calculated this for the
$c_{2,1}=-2$ representation ${\cal R}_{2,1}$
with lowest lying vector $\xi_{2,1}$ at $h=0$ 
and Jordan cell $(\phi_{2,1} = L_{-1} \, \xi_{2,1} \, , \, \psi_{2,1})$ at $h=1$, a representation of
type {\bf B} (see figure \ref{rep_on_border}), and found the following 
first nontrivial nullvector at level $6$ (above the lowest lying vector)
\beq
&& \!\!\!\!\!\!\!\! \Big( m_{1} L_{-1}^{6} + m_{2} L_{-2} L_{-1}^{4} + m_{3} L_{-3} L_{-1}^{3} 
+ (\frac{16}{3}-4 m_{1}+\frac{16}{3} \beta-2 m_{2}) L_{-2}^{2} L_{-1}^{2} \nonumber \\
&& + (-12-12 m_{1}+6 \beta) L_{-4} L_{-1}^{2} + (-\frac{20}{3}-2 m_{3}-16 m_{1}
-\frac{56}{3} \beta-2 m_{2}) L_{-3} L_{-2} L_{-1} \nonumber \\
&& + (\frac{4}{3}-16 m_{1} +\frac{10}{3} \beta-2 m_{2}) L_{-5} L_{-1}  -8  \beta L_{-4} L_{-2} 
+ 6  \beta L_{-3}^{2}  -4  \beta L_{-6} \Big) \; \xi_{2,1}  \nonumber \\
&& \!\!\!\!\!\!\!\!  + \Big( L_{-1}^{5}  -10 L_{-2} L_{-1}^{3} + 6 L_{-3} L_{-1}^{2} + 16 L_{-2}^{2} L_{-1}  
-12 L_{-4} L_{-1}  -8 L_{-3} L_{-2} + 4 L_{-5} \Big) \; \psi_{2,1} \, .
\eeq

This level is indeed the expected one as the Kac table of $c_{2,1}=-2$ gives us
a third nullvector condition for $h=0$ exactly at level $6$
as well as  a corresponding second nullvector condition for $h=1$
at level $5$.
Hence, we confirm the existence of a further nontrivial nullvector at the expected level
in the logarithmic rank $2$ representation ${\cal R}_{2,1}$ derived by
different means in \cite{GK96}.

We notice that up to the overall normalisation of this state the nullvector polynomial
applied to the second Jordan cell state $\psi_{2,1} $ is unique. On the other
hand, the nullvector polynomial on $\xi_{2,1}$, which serves as a correction
to the effects of the indecomposable action on $\psi_{2,1}$, still exhibits
three degrees of freedom. But we know that there is an ordinary nullvector at level $2$
above $h=1$ in the irreducible subre\-presentation whose descendants span a parameter space
of dimension three at level $5$ (above $h=1$). Adding such a descendant of this nullvector
will certainly not alter our equations and, hence, accounts for the additional three
degrees of freedom $m_i$, $i=1,2,3$.

In the same manner one can calculate logarithmic nullvectors in all rank $2$ logarithmic
representations in the $c_{p,1}$ models, limited only by computer power and memory.
We give a second example for a type {\bf B} logarithmic nullvector in appendix \ref{app0}.

%%%%%%%%%

\section{Possible logarithmic nullvectors for $\mathbf{c_{2,3}=0}$\label{section3}}

The $c_{p,1}$ models might be the best-studied logarithmically conformal
models but they still are quite special cases of the general augmented
$c_{p,q}$ models, which we still know much less about.
Hence, an even more exciting question than the above construction of predicted
logarithmic nullvectors surely is whether one can use these techniques
to explore the shapes of the supposedly more complicated logarithmic
representations in general augmented CFTs. In the following we will
attack this question for the augmented model $c_{2,3}=0$ which seems
sufficiently generic to show all the features of general augmented
$c_{p,q}$ models.

\renewcommand{\arraystretch}{0.03}
\begin{table} \caption{Augmented Kac table for $c_{2,3}=0$}
\begin{center}
\begin{tabular}{cc | ccccc}
\multicolumn{7}{c}{\rule[-.6em]{0cm}{1.6em} $\qquad\quad  s$} \\
&\rule[-.6em]{0cm}{1.6em} & $1$ & $2$ & $3$ & $4$ & $5$ \\ \cline{2-7}
& & \multicolumn{5}{c}{} \\
&\rule[-.6em]{0cm}{1.6em} $1$ & $0$ & \cellcolor{gray}{.9}{$\frac{5}{8}$} & $2$ & \cellcolor{gray}{.9}{$\frac{33}{8}$} & $7$ \\
&\rule[-.6em]{0cm}{1.6em} $2$ & $0$ & \cellcolor{gray}{.9}{$\frac{1}{8}$} & $1$ & \cellcolor{gray}{.9}{$\frac {21}{8}$} & $5$ \\
&\rule[-.6em]{0cm}{1.6em} $3$ & \multicolumn{1}{>{\columncolor[gray]{.9}[0.205cm][0.22cm]}c}{$\frac{1}{3}$} 
& \cellcolor{gray}{.7}{$-\frac{1}{24}$} 
& \cellcolor{gray}{.9}{$\frac{1}{3}$} & \cellcolor{gray}{.7}{$\frac {35}{24}$} & \cellcolor{gray}{.9}{$\frac{10}{3}$} \\
$r$ &\rule[-.6em]{0cm}{1.6em} $4$ & $1$ & \cellcolor{gray}{.9}{$\frac{1}{8}$} & $0$ & \cellcolor{gray}{.9}{$\frac{5}{8}$} & $2$ \\
&\rule[-.6em]{0cm}{1.6em} $5$ & $2$ & \cellcolor{gray}{.9}{$\frac{5}{8}$} & $0$ & \cellcolor{gray}{.9}{$\frac{1}{8}$} & $1$ \\
&\rule[-.6em]{0cm}{1.6em} $6$ & \multicolumn{1}{>{\columncolor[gray]{.9}[0.205cm][0.22cm]}c}{$\frac{10}{3}$}
& \cellcolor{gray}{.7}{$\frac{35}{24}$} 
\rule[-.6em]{0cm}{1.6em}& \cellcolor{gray}{.9}{$\frac{1}{3}$} & \cellcolor{gray}{.7}{$-\frac{1}{24}$} & \cellcolor{gray}{.9}{$\frac{1}{3}$} \\
&\rule[-.6em]{0cm}{1.6em} $7$ & $5$ & \cellcolor{gray}{.9}{$\frac {21}{8}$} & $1$ & \cellcolor{gray}{.9}{$\frac{1}{8}$} & $0$ \\
&\rule[-.6em]{0cm}{1.6em} $8$ & $7$ &\cellcolor{gray}{.9}{ $\frac {33}{8}$} & $2$ & \cellcolor{gray}{.9}{$\frac{5}{8}$} & $0$
\end{tabular}
\end{center} \label{Kac0}
\end{table}
\renewcommand{\arraystretch}{1.3}

\subsection{Some facts about augmented Kac tables}
Minimal models manage to extract the smallest possible representation
theory from the Kac table of some central charge $c_{p,q}$ by mapping all weights
to some standard cell using the relations \cite{BPZ84}
\beqn \label{minmod_relations}
h_{r,s} &=& h_{q-r,p-s} \nonumber \\
h_{r,s} &=& h_{r+q,s+p} \qquad \qquad \forall \; 1 \le r < q \quad  1 \le s < p \; .
\eeqn
The only weights which these relations do not relate to anything are the
weights on the border of the Kac table, i.e.\ the weights for $r=q$ or $s=p$ as 
well as their integer multiples. But as these do not appear in any
fusion rule of the other fields in the bulk of the Kac table they are simply ignored.

Augmenting the theory with fields beyond this standard cell
has led to the construction of consistent CFTs containing
representations with non-trivial Jordan blocks, examples of logarithmic CFTs.
These theories can be associated to Kac tables which
comprise the standard cell for larger parameters $p$ and $q$
yielding the same conformal charge; these parameters are usually
odd integer multiples of $p$ and $q$, i.e.\ we deal with the standard cell of 
$c_{n p, n q}$, $n \in (2\N-1)$.
Hence, these theories also contain fields with weights on the border of the
original smaller Kac table as well as fields in the corners, which are the intersections of
the borders.
To give an example, we indicated the borders as areas with lighter shade, the corners
as areas with darker shade in the augmented Kac table of $c_{2,3}$ 
given in table \ref{Kac0}; the bulk consists of the unshaded areas.
All fields of the augmented Kac tables yield independent
representations only subject to the relations \mref{minmod_relations} for the
augmented cell, i.e.\ $p \mapsto n p$, $q \mapsto n q$.
This is the result of the nontrivial fusion of the fields on the border and the corners
with themselves and the fields from the bulk.
Indeed, this fusion behaviour prevents the theory
from just becoming a tensor product of several independent minimal model sectors.

Actually, the only well studied models up to now are contained in the
series $c_{p,1}$, $p=2,3,\dots$ (see e.g. \cite{GK96,GK96b,Flohr95,Flohr96,Flohr01}). 
These models are not generic
in that respect that they do not contain any nontrivial bulk in their
Kac table; they only consist of fields from the border and corners.

The first and easiest example which exhibits a non-empty bulk of the Kac table 
is the augmented $c_{2,3}=0$ model with the  Kac table of $c_{6,9}=0$
which we will explore more closely in the following using our
techniques of logarithmic nullvectors. As mentioned before, 
the corresponding Kac table is given in table \ref{Kac0}.

The fact that we can restrict the representation theory of our CFT to a
finite field content as with the above cells of the Kac table
certainly relies on the enlargement of the symmetry algebra
from a simple Virasoro algebra to a nontrivial \W-algebra, be
it either in the minimal model or the augmented model case.
In the following, however, we will restrict our focus
to the Virasoro algebra and representations
thereof. Still, having the possibility of such an
enlarged \W-algebra in mind, we will mainly focus on representations
connected to weights in the  augmented Kac table cell keeping
in mind that from the point of view of the Virasoro algebra
there is an infinite tower of representations.

%%%

\subsection{Weights on the corners and borders of the augmented 
Kac table of $\mathbf{c_{2,3}=0}$}

We propose that fields associated to weights on the corner and the borders
of the augmented Kac table of $c_{2,3}=0$ are contained in the same types of
representations as the corresponding ones in the $c_{p,1}$ models.

The weights on the corners of the Kac table, $h=-1/24, 35/24$, actually only
appear once modulo the relations \mref{minmod_relations} and accordingly
only exhibit the usual two nullvector conditions.
Hence, there are no new (logarithmic) representations to be expected
besides the ordinary irreducible Virasoro representation
built on groundstates with these weights.
Indeed, these weights give exactly the prelogarithmic fields which
have been shown to be primary and to generate an irreducible representation,
but not to admit an embedding into any larger indecomposable
representation \cite{KL97}.

The weights on the borders of the Kac table actually appear in the
same kind of triplets of two equal conformal weights and one which is
shifted by some positive integer as we know it from the $c_{p,1}$ models (again
modulo the relations \mref{minmod_relations}). The triplets are
$T_1:=\{1/8,1/8,33/8\}$, $T_2:=\{5/8,5/8,21/8\}$ and $T_3:=\{1/3,1/3,10/3\}$.
We also find the same nullvector structure concerning these weights within 
the Kac table as we know it from the corresponding representations
of the $c_{p,1}$ models. Hence,
we have checked the existence of the typical
logarithmic nullvectors for all cases which were accessible
to computer power and memory.

First we have checked for the first logarithmic nullvector
in representation type {\bf A} (see figure \ref{rep_on_border}) and
found the expected ones for all three triplets, on level $8$
for $T_1$,  on level $10$ for $T_2$ as well as on level $9$ for
$T_3$. The result for $T_1$ can be found in appendix \ref{app1}.

A check for the  first logarithmic nullvector
in representation type {\bf B} was only possible for the
triplet $T_2$. In this case, we have a Jordan cell for
$h=21/8$ with a lower lying field at $h=5/8$.
We find the first nontrivial logarithmic nullvector
at level $16$ which seems to be just at the limit of
our current computing power and ability. 
The first logarithmic nullvectors for type {\bf B}
representations corresponding to the other
two triplets are expected at even higher
levels, at $18$ and $20$ for $T_3$ resp.\ $T_1$.

This is indeed compatible with our above proposition
and a nice and nontrivial check for its validity.
The above proposition is also substantiated by the analysis of possible modular
functions of the corresponding \W-algebra \cite{EFN}.

%As e.g.\ the analysis of possible modular functions strongly suggests
%that the representations corresponding to the border of the Kac table in
%$c_{2,3}=0$ behave just the same way as the rank $2$ representations
%in the $c_{p,1}$ models, we will concentrate on possible
%logarithmic representations associated to fields of weights in the
%so-called ``bulk'' of the Kac table.

%%%

\begin{figure}
\centering \leavevmode
\psfig{file=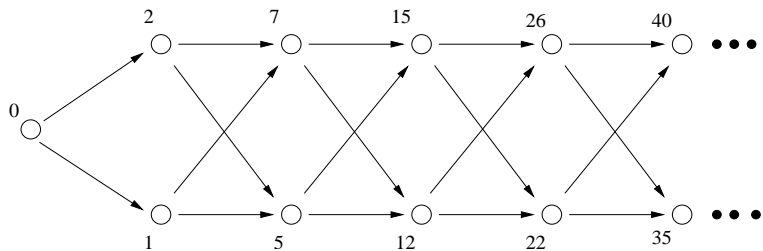,width=10cm}
\caption{Embedding structure for $h=0$ (numbers refer to weights)}
\label{embedding}
\end{figure}

\subsection{Weights in the bulk of the augmented Kac table of $\mathbf{c_{2,3}=0}$}

For possible logarithmic representations corresponding
to weights in the bulk of the Kac table, however,
we do not have any prototypes yet. Hence, we are now going
to explore the main candidates for such representations
and elaborate constraints on their shapes
using our techniques of constructing logarithmic nullvectors.
We notice two main
differences to the situation on the borders.

First of all, the bulk of the augmented Kac table of $c_{2,3}=0$ (see table \ref{Kac0}) 
exhibits an even higher abundance of equal numbers 
(up to integer shift) than in the
$c_{p,1}$ models, which is a clear sign of logarithmic representations there.
Up to the relations \mref{minmod_relations} we actually find
a nonuplet $N = \{0,0,0,1,1,2,2,5,7\}$ of weights which
are equal up to integer shift and which contain one weight with triple 
degeneracy, $h=0$. It does not seem very likely that this nonuplet just splits
into three triplets of the types analysed above.
On the contrary, the analysis of the corresponding modular functions even suggests the
possibility of a rank $3$ logarithmic representation, and certainly
predicts the existence of several
more complicated rank $2$ logarithmic representations 
constructed with weights within this set \cite{EFN}.

On the other hand, we have to notice that the embedding structure for nullvectors
is different in the bulk in contrast to the linear embedding structure
on the border (discussed in \cite{GK96}). 
In the bulk the embedding structure is given by the more generic
two string twisted picture, depicted in figure \ref{embedding},
which can be calculated according to general arguments in \cite{FF82}
or the Virasoro character formula of \cite{Roch85}.\footnote{We
thank A. Nichols for pointing this out to us.}

\begin{figure}
\centering \leavevmode
\psfig{file=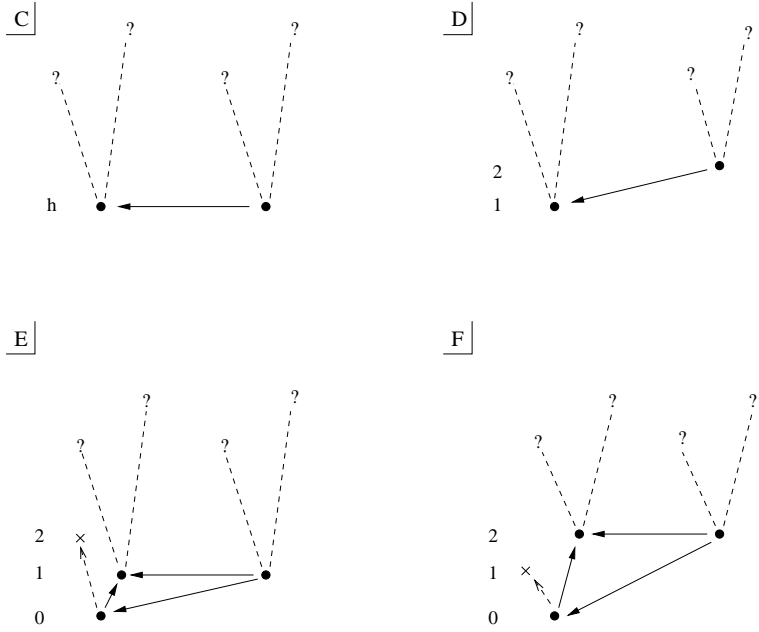,width=10cm}
\caption{Candidates for rank $2$ representations for weights in the bulk of $c=0$}
\label{rep_on_bulk}
\end{figure}
Now, inspecting the nonuplet $N$ of bulk weights 
we expect the usual irreducible re\-presentations to
the integer weights $h=0,1,2,5,7$ as well as rank $2$ representations 
corresponding to Jordan cells at weight $h=0,1,2$.
We have depicted a list of possible candidates
for rank $2$ representations corresponding to these
bulk weights in figure \ref{rep_on_bulk}. These
pictures represent the low lying embedding structure
of these candidates using the same
symbols as in section \ref{section2}. Additionally,
we have indicated the conformal weight on the different
levels to the left of each picture as well as the 
unknown higher embedding structure by question marks (``?'').
We have
checked for the lowest nontrivial logarithmic nullvectors
for all these candidates and summarise the results in table \ref{tab}.

\renewcommand{\arraystretch}{1.0}
\begin{table} \caption{Lowest logarithmic nullvectors of candidates for bulk representations}
\begin{center}
\begin{tabular}{c | c  | c ||  c}
type & lowest weight & rank & level of lowest \\
& & & logarithmic nullvector \\ \hline
C & $0$ & $2$ & $5$ \\
C & $1$ & $2$ & $11$ \\
C & $2$ & $2$ & $10$ \\
C & $0$ & $3$ & $12$ \\
D & $1$ & $2$ & $>14$ \\
E & $0$ & $2$ & $12$ \\
F & $0$ & $2$ & $12$
\end{tabular}
\end{center} \label{tab}
\end{table}

\vspace{0.2cm} \noindent
{\bf Type C.}\hspace{0.1cm}
The calculations for the type {\bf C} representations have been performed
using the methods of section \ref{section1}. For this type we even
managed to calculate one rank $3$ logarithmic nullvector; i.e.\ the first
rank $3$ logarithmic nullvector with lowest weight Jordan cell at $h=0$
appears at level $12$.

\vspace{0.2cm} \noindent
{\bf Type E.}\hspace{0.1cm}
We were able to apply the procedure of section \ref{section2} directly
to the type {\bf E} representation because we do not encounter
any additional nullvector below the level of the Jordan cell
and because we can take $(L_0-h)$ to map $|h;1 \rangle$ 
to a proper singular descendant of $|h-l \rangle$, 
i.e.\ $L_{-1} \, |h-l \rangle$.

\vspace{0.2cm} \noindent
{\bf Type F.}\hspace{0.1cm}
In case of the type {\bf F} representation we actually encounter
an additional nullvector below the level of the Jordan cell.
This can, however, be remedied quite easily.
Due to the lower lying nullvector at level $1$ there is
no state which $L_1$ could map $|h;1 \rangle$ to. But certainly
$L_2$ can take the job to map $|h;1 \rangle$ directly
down to $|h-l \rangle$, a mapping unique up to normalisation.
This yields the new conditions
\beq
L_2 |h;1 \rangle =\beta \: |h-l \rangle &\qquad \quad & L_p |h;1 \rangle = 0 \qquad \; \mbox{for } p = 1 \mbox{ and } p \ge 3 \; .
\eeq
The singular descendant in the Jordan cell is therefore
given by $|h;0 \rangle = L_{-2} \,|h-l \rangle$.

This feature of an additional nullvector below the
level of the Jordan cell clearly shows the novelty of
the bulk representations in contrast to the ones on 
the border described in \cite{GK96};
it arises due to the more intricate embedding structure
for these bulk representations compared to the 
embedding structure for representations on the border.
A generalisation to similar cases with additional
nullvectors on levels lower than the Jordan cell
seems straightforward though more tedious due to the
more complex embedding structure of nullvectors which are
not on the ``nice'' level $1$.

\vspace{0.2cm} \noindent
{\bf Type D.}\hspace{0.1cm}
The case of the type {\bf D} representation is more questionable.
Here, we actually do not have a singular descendant of $|h-l \rangle$
on the level of the Jordan cell, hence no primary state
which $(L_0-h)$ could map $|h;1 \rangle$ to.
A priori it is not clear whether it is necessary for $L_0$
to map $|h;1 \rangle$ to a singular descendant of $|h-l \rangle$.
Hence we took  $(L_0-h)$ to map $|h;1 \rangle$ to the only existing
descendant of $|h-l \rangle$ at level $1$ which is though not singular, 
i.e.\ $L_{-1} \, |h-l \rangle$.
The results of possible logarithmic nullvectors, however, do not 
seem to offer a particular rich structure up to the accessible 
levels (see table \ref{tab}).

\vspace{0.2cm} 
We include the
explicit results for the two cases {\bf E} and {\bf F} in appendix \ref{app2}.
It is quite interesting to inspect e.g.\ the result for type {\bf F}.
Although we actually did not impose the relation $L_{-1} |h-l \rangle = 0$
into the computer programme, the result incorporates such a nullvector
or, to be a bit more cautious, at least the independence of the result
from this particular descendant; indeed, all descendants of $L_{-1} |h-l \rangle$
just appear with free parameters. This corresponds to the additional
freedom due to lower nullvectors in the irreducible subrepresentation
discussed in the end of section \ref{section2}.
On the other hand, the second singular vector on $|h-l \rangle$ on level $2$ 
does not
pop up in the same manner as a possible nullvector in this result; 
rather, the result is consistent with our ansatz where we actually
impose that the level $2$ singular vector on $|h-l \rangle$
is not a nullvector for the whole rank $2$ representation.
Hence, we take the first observation of the independence
from $L_{-1} |h-l \rangle$ as a strong hint that a representation
where both these singular descendants are not null in the whole logarithmic
representation is not favoured by our calculations.

%%%%%%%%%

\section{Conclusion and outlook}

In the preceding sections we have elaborated the procedure how
to calculate generic logarithmic nullvectors. This generalises former
computations, which assumed that all fields in Jordan cells were (logarithmic)
quasi-primary. We have calculated
several examples of such nullvectors in rank $2$ indecomposable
representations, for two $c_{p,1}$ models as well as for
the more generic augmented $c_{2,3} = 0$ model.

Although these logarithmic nullvectors do not enable us
to write down the whole indecomposable representations
and hence to decide which of these representations are
realised in our model,
they nevertheless provide us with severe constraints
about the number of states on the lower levels
for all inspected candidates. As discussed for 
re\-pre\-sentation type {\bf F} one can even use
the calculated logarithmic nullvector to give
strong arguments for the existence of
singular vectors of the irreducible subrepresentation
as nullvectors of the whole indecomposable representation.

On the other hand, we would like to stress again
that we can regard the augmented $c_{2,3} = 0$ model
as a prototype for the general augmented $c_{p,q}$ models.
As for the series of $c_{p,1}$ models where one encounters
the same structure of singlets on the corners and triplets
on the borders throughout the series, 
we find the same structures which we have
described in section \ref{section3} for all
$c_{p,q}$ models: singlets on the corners, triplets
on the borders and nonuplets in the bulk --- only with
a larger variety for larger models. And, most importantly, 
we find the same kind of relations between the levels of possible
nullvectors and the differences between the weights
in the n-plets as the ones described for $c_{2,3} = 0$ 
in section \ref{section3}.

Concerning the classification of nullvectors in logarithmic CFT
\cite{Flohr97,Mog00,Mog02}, 
we observe that the generic nullvectors of rank $2$
exist precisely at the levels one would guess from the Kac table, where
the level has to be counted from the conformal weight of the Jordan cell.
Within the bulk, it seems possible that even rank $3$ generic nullvectors 
exist. However, we have only been able to calculate one example
of a rank $3$ logarithmic nullvector in the $h=0$ sector of the 
$c_{2,3}$ model so far.

Thus, we made use of the information 
about logarithmic nullvectors to give
a rough picture about how the embedding structure of rank $2$
logarithmic representations might look like
for the up to now almost unexplored
augmented $c_{p,q}$ models.

Certainly one can now use these explicitely
constructed rank $2$ nullvectors to calculate corresponding
correlation functions and hence analyse the physical dynamics
of these theories. This would have to proceed along the lines
indicated in \cite{Flohr00}.

On the other hand, one seems to be very close to finally 
pinpoint the full structure of the rank $2$ representations
in the $c_{2,3}$ model and, hence, in all $c_{p,q}$ models.
In order to achieve this it seems necessary to bring together
the constraints on possible rank $2$ representations calculated
in this paper with the knowledge about the representation
of the modular group corresponding to the inherent
enlarged \W-algebra. Indeed, this larger symmetry algebra
makes it possible to uniquely fix any representation of $c_{2,3}$
if one knows the multiplicities of states up to level $7$.
But it is still not clear how to combine Virasoro representations
to full \W-representations. These considerations are subject
to ongoing research \cite{EFN}. Furthermore, the construction of these
representations will supposedly also settle the long standing
puzzle around the structure of the vacuum character in
the augmented $c_{2,3} = 0$ model (see e.g.\ \cite{FM05}).
On the other hand, it does not seem too far out of reach to construct generic 
\W-nullvectors
along the lines set out in this work. The knowledge of such nullvectors
would enable us to prove rationality, or at least $C_2$-cofiniteness, for
augmented $c_{p,q}$ models using the approach of \cite{CF05}.

After completion of this work the article \cite{FGST05} appeared
which proves the equivalence between the $c_{2,1}$ model as a vertex
operator algebra and a corresponding quantum group and
conjectures a similar result for general $c_{p,1}$ models.
This manifestation of the Kazhdan--Lusztig correspondence
as an equivalence for the case of the $c_{p,1}$ models leads
to a highly interesting insight in the quantum group structure
of these conformal field theories and might facilitate the
classification of representations in these models.
We hope that results as the ones about logarithmic nullvectors 
in this article will finally lead to a generalisation of \cite{FGST05} to general
$c_{p,q}$ models, although regarding our still very small knowledge
about these models this seems still a long way to go.

%%%%%%%%%%%%%%%%%%%%%%%%%%%%%%%%%%%%%%%%%%%%%

\vspace{0.7cm}
\noindent
{\bf Acknowledgements.}
We would like to thank Werner Nahm and Alexander Nichols
for useful discussions.
This work was supported in part by 
the DFG Schwerpunktprogramm no.\ 1096 ``Stringtheorie''.
HE acknowledges support by the Studienstiftung des Deutschen Volkes.
MF is also partially supported by  the European Union network 
HPRN-CT-2002-00325 (EUCLID).

%%%%%%%%%%%%%%%%%%%%%%%%%%%%%%%%%%%%%%%%%%%%%

\begin{appendix}

\section{Implementation of the logarithmic nullvector calculation on 
the computer\label{computer}}

The calculations performed for this article have been implemented
in C++ using the computer algebra package GiNaC \cite{Ginac}.
We constructed new classes for the algebraic objects fields,
fieldmodes, products of fieldmodes as well as descendant fields.
The implemented Virasoro relations (as well as possibly further
commutation relations) are used for direct simplification
of descendant fields towards the normal ordered standard form
as soon as these are constructed. It is important to note that
within the procedure the application of modes to the field has 
priority to the commutation of modes in order to reduce the blow-up of the
number of terms within the calculation.

The calculation of matrix elements is performed in two steps.
First all fieldmodes within the matrix elements are used to construct a
descendant state on the ket-state which is then automatically
simplified (see above). Then the correct coefficients
are picked using the properties of the bra-state as well as 
the Shapovalov form.

The main property which has to be implemented into the fields,
besides their conformal weight (and possibly fermion number), is
their behaviour under the action of nonnegative Virasoro modes.
We picked conformal primarity as the standard and implemented
deviations from that in a list which is pointed to by a member
of each instance. E.g.\ for calculations of representation
type {\bf B} (see figure \ref{rep_on_border}) we had to implement
the indecomposable $L_0$ action as well as the non-vanishing
$L_1$ action which maps $|h;1 \rangle$ to the unique 
level $(l-1)$ descendant of $|h-l \rangle$ with properties 
\mref{unique}.

%%%%%%%%%

\section{An explicit nullvector for $\mathbf{c_{3,1} = -7}$\label{app0}}
As a further example for a type {\bf B} rank 2 logarithmic nullvector (see figure \ref{rep_on_border})
we give the respective nullvector with lowest lying vector at $h=-1/4$ and Jordan cell
at $h=7/4$ in the augmented model of $c_{3,1} = -7$ which appears at level $10$.
For the sake of reasonable brevity we have set the overall normalisation to $1$
and also eliminated any freedom due to the existence of lower
nullvectors in the irreducible subrepresentation by setting any further free parameter to $0$;
the parameter $\beta$ certainly still remains as it is a parameter of the representation as introduced in
section \ref{section2}:
\beq 
\lb \Big( (\frac{7168}{27}-512 \beta) L_{-3} L_{-2}^{2} L_{-1}^{3} + (-\frac{1280}{81}+\frac{256}{9} \beta) L_{-5} L_{-2} L_{-1}^{3} + (-\frac{21376}{27}-\frac{128}{3} \beta) L_{-4} L_{-3} L_{-1}^{3} \\
\lb + (\frac{8552}{27}+\frac{3280}{9} \beta) L_{-7} L_{-1}^{3} + (-\frac{4096}{27}+\frac{512}{3} \beta) L_{-2}^{4} L_{-1}^{2} + (\frac{67840}{81}+\frac{256}{9} \beta) L_{-4} L_{-2}^{2} L_{-1}^{2} \\
\lb + (-\frac{18112}{81}+\frac{8192}{9} \beta) L_{-3}^{2} L_{-2} L_{-1}^{2} + (\frac{30080}{81}-\frac{11008}{9} \beta) L_{-6} L_{-2} L_{-1}^{2} + (8+48 \beta) L_{-5} L_{-3} L_{-1}^{2}  \\
\lb + (-\frac{3968}{27}-\frac{736}{3} \beta) L_{-4}^{2} L_{-1}^{2} + (\frac{24752}{81}-\frac{2992}{9} \beta) L_{-8} L_{-1}^{2} + (-\frac{7168}{81}+\frac{512}{3} \beta) L_{-3} L_{-2}^{3} L_{-1} \\
\lb + (\frac{1280}{243}+\frac{2048}{27} \beta) L_{-5} L_{-2}^{2} L_{-1} + (\frac{21376}{81}+\frac{3968}{9} \beta) L_{-4} L_{-3} L_{-2} L_{-1} -\frac{3008}{9}  \beta L_{-3}^{3} L_{-1} \\ 
\lb + (-\frac{8552}{81}-496 \beta) L_{-7} L_{-2} L_{-1} + \frac{2720}{3}  \beta L_{-6} L_{-3} L_{-1}  -208  \beta L_{-5} L_{-4} L_{-1} + \frac{5564}{9}  \beta L_{-9} L_{-1} \\
\lb + (\frac{4096}{81}-\frac{512}{9} \beta) L_{-2}^{5} + (-\frac{67840}{243}-\frac{6400}{27} \beta) L_{-4} L_{-2}^{3} + (-\frac{3392}{243}+\frac{1024}{27} \beta) L_{-3}^{2} L_{-2}^{2} \\
\lb + (-\frac{30080}{243}+\frac{11776}{27} \beta) L_{-6} L_{-2}^{2} + (-\frac{42376}{243}-\frac{496}{27} \beta) L_{-5} L_{-3} L_{-2} + (\frac{3968}{81}+\frac{4576}{9} \beta) L_{-4}^{2} L_{-2} \\
\lb + (-\frac{110768}{243}+\frac{15904}{27} \beta) L_{-8} L_{-2} + (\frac{21376}{81}-\frac{832}{9} \beta) L_{-4} L_{-3}^{2} + (-\frac{30056}{81}+288 \beta) L_{-7} L_{-3} \\
\lb + \frac{448}{3}  \beta L_{-6} L_{-4} + (\frac{1280}{243}+\frac{1580}{27} \beta) L_{-5}^{2} + (-\frac{28672}{27}+2128 \beta) L_{-10} \Big) \; \cet{h-l} \\
\lb +  \Big( L_{-1}^{8}  -28 L_{-2} L_{-1}^{6} + 28 L_{-3} L_{-1}^{5} + \frac{658}{3} L_{-2}^{2} L_{-1}^{4}  -132 L_{-4} L_{-1}^{4}  -\frac{1048}{3} L_{-3} L_{-2} L_{-1}^{3} \\
\lb + \frac{976}{9} L_{-5} L_{-1}^{3}  -\frac{12916}{27} L_{-2}^{3} L_{-1}^{2} + \frac{8344}{9} L_{-4} L_{-2} L_{-1}^{2} + \frac{1604}{9} L_{-3}^{2} L_{-1}^{2} + 240 L_{-6} L_{-1}^{2} \\
\lb + 668 L_{-3} L_{-2}^{2} L_{-1}  -\frac{1312}{3} L_{-5} L_{-2} L_{-1}  -576 L_{-4} L_{-3} L_{-1}  -186 L_{-7} L_{-1}   -\frac{1708}{3} L_{-4} L_{-2}^{2}  \\
\lb + \frac{1225}{9} L_{-2}^{4} -\frac{728}{3} L_{-3}^{2} L_{-2}  -224 L_{-6} L_{-2} + \frac{602}{3} L_{-5} L_{-3} + 252 L_{-4}^{2}  -476 L_{-8} \Big) \; \cet{h;1}
\eeq

%%%%%%%%%

\section{Explicit nullvectors on the border of $\mathbf{c_{2,3} = 0}$\label{app1}}

In the following we give the explicit form of the nullvector 
of type {\bf A} for the triplet $T_1$, which has a Jordan cell at lowest 
weight $h = 1/8$ and appears at level $8$. We have set the overall
normalisation to $1$ in this expression; any further free parameters,
however, appear as calculated (noted as $m_i$). We also note again
that $\beta$ is not a free parameter of the logarithmic nullvector
calculation but a parameter of the representation as introduced in
section \ref{section2}:
\beq
\lb \Big( m_{21} L_{-1}^{8} + m_{20} L_{-2} L_{-1}^{6} + m_{19} L_{-3} L_{-1}^{5} + m_{18} L_{-2}^{2} L_{-1}^{4} + m_{17} L_{-4} L_{-1}^{4} \\
\lb + (\frac{64}{9}-\frac{13}{3} m_{19}-\frac{278}{3} m_{21}-7 m_{20}) L_{-3} L_{-2} L_{-1}^{3} + (\frac{616}{9}-\frac{698}{9} m_{21}-7 m_{20}) L_{-5} L_{-1}^{3} \\
\lb + (-\frac{4069}{54} m_{21}-\frac{13}{3} m_{18}-\frac{217}{12} m_{20}) L_{-2}^{3} L_{-1}^{2} + (\frac{2870}{9}+15 m_{21}-5 m_{20}) L_{-6} L_{-1}^{2} \\
\lb + (\frac{752}{9}-\frac{1475}{9} m_{21}-\frac{5}{2} m_{20}-\frac{13}{3} m_{17}) L_{-4} L_{-2} L_{-1}^{2} + (-\frac{640}{9}-\frac{35}{12} m_{21}-\frac{1}{2} m_{17}) L_{-4}^{2} \\
\lb + (-\frac{2164}{9}-\frac{8}{3} m_{19}-\frac{301}{9} m_{21}) L_{-3}^{2} L_{-1}^{2} + (\frac{5425}{432} m_{21}+\frac{25}{36} m_{18}+\frac{325}{108} m_{20}) L_{-2}^{4} \\
\lb + (-540+142 m_{21}+\frac{10}{3} m_{18}+\frac{461}{18} m_{20}) L_{-5} L_{-2} L_{-1} + (\frac{1888}{27}+\frac{25}{36} m_{19}+\frac{955}{18} m_{21}+ \\
\lb \frac{5}{3} m_{18}+10 m_{20}) L_{-3} L_{-2}^{2} L_{-1} + (\frac{2768}{9}+\frac{17}{6} m_{19}+143 m_{21}+\frac{5}{3} m_{17}) L_{-4} L_{-3} L_{-1} \\
\lb+ (-\frac{614}{3}+\frac{17}{6} m_{19}+\frac{14059}{36} m_{21}+5 m_{18}+\frac{281}{6} m_{20}) L_{-7} L_{-1} + (-\frac{104}{9}+\frac{1087}{36} m_{21} \\
\lb -\frac{1}{2} m_{18} +\frac{11}{18} m_{20}+\frac{25}{36} m_{17}) L_{-4} L_{-2}^{2} + (\frac{266}{27}+\frac{25}{18} m_{19}+\frac{250}{9} m_{21}+\frac{25}{18} m_{20}) L_{-3}^{2} L_{-2} \\
\lb + (-\frac{43}{3}+\frac{131}{18} m_{21}-2 m_{18}-\frac{19}{3} m_{20}) L_{-6} L_{-2} + (\frac{2482}{27}-\frac{29}{36} m_{19}+\frac{131}{4} m_{21} \\
\lb +\frac{25}{9} m_{20}) L_{-5} L_{-3} + (\frac{1220}{27}-\frac{29}{18} m_{19}-\frac{3985}{18} m_{21}-4 m_{18}-\frac{247}{9} m_{20}) L_{-8} \Big) \; \cet{h-l} \\
\lb +  \Big( L_{-1}^{8}  -\frac{58}{3} L_{-2} L_{-1}^{6} + \frac{70}{3} L_{-3} L_{-1}^{5} + \frac{1547}{18} L_{-2}^{2} L_{-1}^{4}  -\frac{347}{3} L_{-4} L_{-1}^{4} -\frac{526}{9} L_{-3} L_{-2} L_{-1}^{3} \\
\lb + \frac{520}{9} L_{-5} L_{-1}^{3} -\frac{589}{6} L_{-2}^{3} L_{-1}^{2} + \frac{1157}{3} L_{-4} L_{-2} L_{-1}^{2}  -\frac{287}{3} L_{-3}^{2} L_{-1}^{2} + \frac{335}{3} L_{-6} L_{-1}^{2} \\
\lb + \frac{115}{6} L_{-3} L_{-2}^{2} L_{-1} -\frac{200}{3} L_{-5} L_{-2} L_{-1} + \frac{49}{3} L_{-4} L_{-3} L_{-1}  -\frac{229}{12} L_{-7} L_{-1} + \frac{225}{16} L_{-2}^{4} \\
\lb -\frac{1259}{12} L_{-4} L_{-2}^{2} + \frac{100}{3} L_{-3}^{2} L_{-2}  -\frac{253}{6} L_{-6} L_{-2}  -\frac{159}{4} L_{-5} L_{-3} + \frac{659}{12} L_{-4}^{2}  -\frac{433}{6} L_{-8} \Big) \; \cet{h;1}
\eeq

We also give the type {\bf B} logarithmic nullvector for the triplet $T_2$,
which has a Jordan cell at $h=21/8$, a lowest weight at $h=5/8$ and
appears at level $16$.
As this expression is very lengthy we have again 
set the overall normalisation to $1$
and also eliminated any further freedom by setting any further free parameter to $0$:
\beq
\lb \Big( (-\frac{149120}{27}-\frac{11648}{9} \beta) L_{-3} L_{-2}^{2} L_{-1}^{9} + (-\frac{6100784}{81}-\frac{4996016}{405} \beta) L_{-5} L_{-2} L_{-1}^{9} + (-\frac{564736}{27}- \\
\lb \frac{2531584}{135} \beta) L_{-4} L_{-3} L_{-1}^{9} + (\frac{1492754}{27}-\frac{761422}{135} \beta) L_{-7} L_{-1}^{9} + (\frac{58240}{27}+\frac{11648}{27} \beta) L_{-2}^{4} L_{-1}^{8} \\
\lb + (\frac{10401664}{81}+\frac{9886336}{405} \beta) L_{-4} L_{-2}^{2} L_{-1}^{8} + (\frac{6547352}{81}+\frac{13290968}{405} \beta) L_{-3}^{2} L_{-2} L_{-1}^{8} + (\frac{11199188}{81} \\
\lb+\frac{11666036}{405} \beta) L_{-6} L_{-2} L_{-1}^{8} + (\frac{11824162}{81}+\frac{16560418}{405} \beta) L_{-5} L_{-3} L_{-1}^{8} + (-\frac{2163322}{27} \\
\lb +\frac{18004358}{135} \beta) L_{-4}^{2} L_{-1}^{8} + (\frac{25195051}{81}-\frac{19264057}{81} \beta) L_{-8} L_{-1}^{8} + (-\frac{298240}{27}-\frac{55552}{27} \beta) L_{-3} L_{-2}^{3} L_{-1}^{7} \\
\lb + (\frac{389529760}{243}+\frac{72031904}{243} \beta) L_{-5} L_{-2}^{2} L_{-1}^{7} + (-\frac{43345088}{81}-\frac{4798528}{81} \beta) L_{-4} L_{-3} L_{-2} L_{-1}^{7} \\
\lb + (-\frac{161243044}{81} +\frac{264057116}{405} \beta) L_{-7} L_{-2} L_{-1}^{7} + (-\frac{3836560}{27}-\frac{1439984}{15} \beta) L_{-3}^{3} L_{-1}^{7} + (-\frac{39224696}{81} \\
\lb +\frac{79328632}{405} \beta) L_{-6} L_{-3} L_{-1}^{7} + (-\frac{30255232}{81}-\frac{420183616}{405} \beta) L_{-5} L_{-4} L_{-1}^{7} \\
\lb + (-\frac{115606213}{81} +\frac{317139341}{405} \beta) L_{-9} L_{-1}^{7} + (-\frac{546560}{81}-\frac{109312}{81} \beta) L_{-2}^{5} L_{-1}^{6} \\
\lb + (-\frac{504068096}{243}-\frac{414460928}{1215} \beta)  L_{-4} L_{-2}^{3} L_{-1}^{6} + (-\frac{32395888}{243}-\frac{221287024}{1215} \beta) L_{-3}^{2} L_{-2}^{2} L_{-1}^{6} \\ 
\lb + (-\frac{688257640}{243}-\frac{1894942504}{1215} \beta) L_{-6} L_{-2}^{2} L_{-1}^{6} + (-\frac{171246124}{81}-\frac{23791292}{81} \beta) L_{-5} L_{-3} L_{-2} L_{-1}^{6} \\ 
\lb + (\frac{436742516}{81}-\frac{285590284}{405} \beta) L_{-4}^{2} L_{-2} L_{-1}^{6} + (-\frac{1311363362}{243}+\frac{618446318}{243} \beta) L_{-8} L_{-2} L_{-1}^{6} + \\
\lb (-\frac{14349776}{81}+\frac{87406928}{81} \beta)  L_{-4} L_{-3}^{2} L_{-1}^{6} + (\frac{675988264}{81}-\frac{1346752232}{405} \beta) L_{-7} L_{-3} L_{-1}^{6} \\
\lb + (\frac{130901740}{27}+\frac{530789812}{135} \beta) L_{-6} L_{-4} L_{-1}^{6} + (\frac{57193417}{243}+\frac{323247827}{243} \beta) L_{-5}^{2} L_{-1}^{6} \\
\lb + (\frac{377046397}{27}-\frac{82026941}{135} \beta) L_{-10} L_{-1}^{6} + (\frac{24149440}{81}+\frac{478912}{9} \beta) L_{-3} L_{-2}^{4} L_{-1}^{5} \\
\lb + (-\frac{6472052776}{729}-\frac{1153026056}{729} \beta) L_{-5} L_{-2}^{3} L_{-1}^{5} + (\frac{491223904}{81}+\frac{724362016}{405} \beta) L_{-4} L_{-3} L_{-2}^{2} L_{-1}^{5} \\
\lb + (\frac{1289170697}{81}-\frac{2296000231}{405} \beta) L_{-7} L_{-2}^{2} L_{-1}^{5} + (-\frac{247531144}{243}-\frac{33723704}{243} \beta) L_{-3}^{3} L_{-2} L_{-1}^{5} \\
\lb + (\frac{74820020}{9}+\frac{37930060}{9} \beta) L_{-6} L_{-3} L_{-2} L_{-1}^{5} + (-\frac{349441376}{243}+\frac{8040650368}{1215} \beta) L_{-5} L_{-4} L_{-2} L_{-1}^{5} \\
\lb + (\frac{289876117}{18}-\frac{908160193}{162} \beta) L_{-9} L_{-2} L_{-1}^{5} + (-\frac{899217364}{243}-\frac{1223108068}{1215} \beta) L_{-5} L_{-3}^{2} L_{-1}^{5} \\
\lb + (\frac{54747340}{9}-\frac{21043364}{9} \beta) L_{-4}^{2} L_{-3} L_{-1}^{5} + (-\frac{3343815314}{243}+\frac{1020421910}{243} \beta) L_{-8} L_{-3} L_{-1}^{5} \\
\lb + (-\frac{4106356768}{81}+\frac{423802096}{81} \beta) L_{-7} L_{-4} L_{-1}^{5} + (\frac{6087567380}{243}-\frac{1708569956}{243} \beta) L_{-6} L_{-5} L_{-1}^{5} \\
\lb + (\frac{1975182488}{243}-\frac{5191641784}{1215} \beta) L_{-4} L_{-3}^{2} L_{-2} L_{-1}^{4} + (-\frac{10951360}{243}-\frac{2190272}{243} \beta) L_{-2}^{6} L_{-1}^{4} \\
\lb + (\frac{6439826624}{729}+\frac{4398126656}{3645} \beta) L_{-4} L_{-2}^{4} L_{-1}^{4} + (-\frac{805947068}{729}+\frac{946244068}{3645} \beta) L_{-3}^{2} L_{-2}^{3} L_{-1}^{4} \\
\lb + (\frac{9716768158}{729}+\frac{39829789006}{3645} \beta) L_{-6} L_{-2}^{3} L_{-1}^{4} + (\frac{1492444673}{243}-\frac{3094966463}{1215} \beta) L_{-5} L_{-3} L_{-2}^{2} L_{-1}^{4} \\
\lb + (-\frac{10110621079}{243}-\frac{300543275}{243} \beta) L_{-4}^{2} L_{-2}^{2} L_{-1}^{4} + (\frac{12461262967}{486}-\frac{7049560643}{810} \beta) L_{-8} L_{-2}^{2} L_{-1}^{4} \\
\lb + (-\frac{23944200971}{486}-\frac{1603601245}{486} \beta) L_{-11} L_{-1}^{5} + (-\frac{11528458402}{243}+\frac{25042759046}{1215} \beta) L_{-7} L_{-3} L_{-2} L_{-1}^{4} \\
\lb + (-\frac{19090179110}{243}-\frac{62902001258}{1215} \beta) L_{-6} L_{-4} L_{-2} L_{-1}^{4} + (\frac{9182523901}{1458}-\frac{9713083541}{7290} \beta) L_{-5}^{2} L_{-2} L_{-1}^{4} \\
\lb + (-\frac{65932612501}{486}-\frac{3429904271}{486} \beta) L_{-10} L_{-2} L_{-1}^{4} + (\frac{383243984}{243}+\frac{323521072}{243} \beta) L_{-3}^{4} L_{-1}^{4} \\
\lb + (\frac{55495736}{27} -\frac{829672184}{81} \beta) L_{-6} L_{-3}^{2} L_{-1}^{4} + (\frac{1588949072}{243}+\frac{24002270192}{1215} \beta) L_{-5} L_{-4} L_{-3} L_{-1}^{4} \\
\lb + (\frac{957145103}{243}  -\frac{23998651177}{1215} \beta) L_{-9} L_{-3} L_{-1}^{4} + (-\frac{44315552}{27}-\frac{171541792}{27} \beta) L_{-4}^{3} L_{-1}^{4} \\
\lb + (\frac{4968093016}{243}+\frac{5395525736}{243} \beta) L_{-8} L_{-4} L_{-1}^{4} + (\frac{2882901476}{243}-\frac{33207042922}{1215} \beta) L_{-7} L_{-5} L_{-1}^{4} \\
\lb + (-\frac{6761458844}{243}+\frac{3531609476}{243} \beta) L_{-6}^{2} L_{-1}^{4} + (-\frac{5936082307}{243}-\frac{19072395649}{1215} \beta) L_{-12} L_{-1}^{4} \\
\lb + (-\frac{260692160}{729}+\frac{1981504}{81} \beta) L_{-3} L_{-2}^{5} L_{-1}^{3} + (\frac{25860528968}{2187}+\frac{13853999912}{10935} \beta) L_{-5} L_{-2}^{4} L_{-1}^{3} \\
\lb + (-\frac{115094416}{9}-\frac{397874864}{81} \beta) L_{-4} L_{-3} L_{-2}^{3} L_{-1}^{3} + (-\frac{9183968675}{243}+\frac{60366559}{5} \beta) L_{-7} L_{-2}^{3} L_{-1}^{3} \\
\lb + (\frac{22446028}{9}+\frac{381549548}{243} \beta) L_{-3}^{3} L_{-2}^{2} L_{-1}^{3} + (-\frac{1857923638}{243}-\frac{26692922234}{1215} \beta) L_{-6} L_{-3} L_{-2}^{2} L_{-1}^{3} \\
\lb + (\frac{18483518080}{729}-\frac{1635799888}{729} \beta) L_{-5} L_{-4} L_{-2}^{2} L_{-1}^{3} + (-\frac{106944789343}{2916}+\frac{123521710391}{14580} \beta) L_{-9} L_{-2}^{2} L_{-1}^{3} \\
\lb + (\frac{10262941796}{729}+\frac{4345875956}{3645} \beta) L_{-5} L_{-3}^{2} L_{-2} L_{-1}^{3} + (-\frac{2914640996}{81}+\frac{5112097436}{405} \beta) L_{-4}^{2} L_{-3} L_{-2} L_{-1}^{3} \\
\lb + (\frac{41031702454}{729}-\frac{71201971634}{3645} \beta) L_{-8} L_{-3} L_{-2} L_{-1}^{3} + (\frac{81698877104}{243}-\frac{40731621952}{1215} \beta) L_{-7} L_{-4} L_{-2} L_{-1}^{3} \\
\lb + (-\frac{114847983652}{729}+\frac{184898225636}{3645} \beta) L_{-6} L_{-5} L_{-2} L_{-1}^{3} + (\frac{464402212315}{1458}+\frac{154322577601}{7290} \beta) L_{-11} L_{-2} L_{-1}^{3} \\
\lb + (\frac{15314752}{3}-\frac{1668550912}{405} \beta) L_{-4} L_{-3}^{3} L_{-1}^{3} + (-\frac{5194154518}{81}+\frac{2708172854}{135} \beta) L_{-7} L_{-3}^{2} L_{-1}^{3} \\
\lb + (\frac{2137814356}{81}+\frac{221522588}{81} \beta) L_{-6} L_{-4} L_{-3} L_{-1}^{3} + (\frac{1006401133}{729}-\frac{51304031429}{3645} \beta) L_{-5}^{2} L_{-3} L_{-1}^{3} \\
\lb + (-\frac{418142455}{9}+\frac{10153248079}{405} \beta) L_{-10} L_{-3} L_{-1}^{3} + (-\frac{913003520}{81}+\frac{3109906496}{405} \beta) L_{-5} L_{-4}^{2} L_{-1}^{3} \\
\lb + (\frac{3782161967}{81}+\frac{14793947}{135} \beta) L_{-9} L_{-4} L_{-1}^{3} + (-\frac{10430827276}{243}+\frac{697278628}{405} \beta) L_{-8} L_{-5} L_{-1}^{3} \\
\lb + (\frac{4093892420}{81}-\frac{5063706352}{405} \beta) L_{-7} L_{-6} L_{-1}^{3} + (\frac{2816698331}{108}+\frac{14750597275}{324} \beta) L_{-13} L_{-1}^{3} \\
\lb + (-\frac{3876174992}{243}-\frac{71245077136}{1215} \beta) L_{-5} L_{-4} L_{-3} L_{-2} L_{-1}^{2} + (-\frac{15573821056}{2187}-\frac{5099410048}{10935} \beta) L_{-4} L_{-2}^{5} L_{-1}^{2} \\
\lb + (-\frac{240207548}{2187}-\frac{8494313084}{10935} \beta) L_{-3}^{2} L_{-2}^{4} L_{-1}^{2} + (-\frac{47436984506}{2187}-\frac{39860881954}{2187} \beta) L_{-6} L_{-2}^{4} L_{-1}^{2} \\
\lb + (\frac{47109428591}{2187}+\frac{184063255247}{10935} \beta) L_{-5} L_{-3} L_{-2}^{3} L_{-1}^{2} + (\frac{34161020861}{729}-\frac{4963416739}{3645} \beta) L_{-4}^{2} L_{-2}^{3} L_{-1}^{2} \\
\lb + (-\frac{5763567647}{1458}+\frac{2691154249}{162} \beta) L_{-8} L_{-2}^{3} L_{-1}^{2} + (-\frac{13695268444}{729}+\frac{7011515372}{3645} \beta) L_{-4} L_{-3}^{2} L_{-2}^{2} L_{-1}^{2} \\
\lb + (\frac{21517454756}{729}-\frac{116959711288}{3645} \beta) L_{-7} L_{-3} L_{-2}^{2} L_{-1}^{2} + (\frac{196664238547}{729}+\frac{542209846357}{3645} \beta) L_{-6} L_{-4} L_{-2}^{2} L_{-1}^{2}  \\
\lb + (-\frac{238400926535}{2916}-\frac{104126564893}{2916} \beta) L_{-5}^{2} L_{-2}^{2} L_{-1}^{2} + (\frac{855093533125}{2916}+\frac{606232113211}{14580} \beta) L_{-10} L_{-2}^{2} L_{-1}^{2}  \\
\lb + (\frac{1639307648}{729}-\frac{5639533168}{3645} \beta) L_{-3}^{4} L_{-2} L_{-1}^{2} + (-\frac{16056755168}{243}+\frac{3324232744}{243} \beta) L_{-6} L_{-3}^{2} L_{-2} L_{-1}^{2}  \\
\lb + (\frac{9611840}{729}+\frac{1922368}{729} \beta) L_{-2}^{7} L_{-1}^{2} + (-\frac{10283821591}{1458}+\frac{377475370451}{7290} \beta) L_{-9} L_{-3} L_{-2} L_{-1}^{2}  \\
\lb + (\frac{520593056}{81}+\frac{54200032}{3} \beta) L_{-4}^{3} L_{-2} L_{-1}^{2} + (-\frac{119998885100}{729}-\frac{62785077796}{729} \beta) L_{-8} L_{-4} L_{-2} L_{-1}^{2}  \\
\lb + (\frac{8363078144}{729}+\frac{380846363054}{3645} \beta) L_{-7} L_{-5} L_{-2} L_{-1}^{2} + (\frac{82219813000}{729}-\frac{29138179912}{729} \beta) L_{-6}^{2} L_{-2} L_{-1}^{2}  \\
\lb + (\frac{31541761250}{729}+\frac{97639549658}{3645} \beta) L_{-12} L_{-2} L_{-1}^{2} + (\frac{3724220342}{729}+\frac{16033283606}{3645} \beta) L_{-5} L_{-3}^{3} L_{-1}^{2}  \\
\lb + (-\frac{676716122}{81}-\frac{135149054}{135} \beta) L_{-4}^{2} L_{-3}^{2} L_{-1}^{2} + (\frac{39555209293}{729}+\frac{95016781}{3645} \beta) L_{-8} L_{-3}^{2} L_{-1}^{2}  \\
\lb + (\frac{11257024304}{243}+\frac{15187070696}{1215} \beta) L_{-7} L_{-4} L_{-3} L_{-1}^{2} + (-\frac{13835770486}{729}-\frac{20728972126}{3645} \beta) L_{-6} L_{-5} L_{-3} L_{-1}^{2}  \\
\lb + (\frac{161403296071}{1458}+\frac{185298212569}{7290} \beta) L_{-11} L_{-3} L_{-1}^{2} + (-\frac{1353700318}{9}-\frac{5011610942}{81} \beta) L_{-6} L_{-4}^{2} L_{-1}^{2}  \\
\lb + (\frac{39014405471}{729}+\frac{81704195681}{3645} \beta) L_{-5}^{2} L_{-4} L_{-1}^{2} + (-\frac{1251898711}{27}-\frac{3014837587}{405} \beta) L_{-10} L_{-4} L_{-1}^{2}  \\
\lb + (-\frac{54758596976}{729}-\frac{189462959732}{3645} \beta) L_{-9} L_{-5} L_{-1}^{2} + (\frac{26091481631}{243}+\frac{81857918359}{1215} \beta) L_{-8} L_{-6} L_{-1}^{2}  \\
\lb + (\frac{41764853429}{243}-\frac{59364317437}{1215} \beta) L_{-6} L_{-5} L_{-2}^{2} L_{-1} + (\frac{519772644341}{729}+\frac{536982902849}{3645} \beta) L_{-14} L_{-1}^{2} \\
\lb  + (\frac{16951960}{243}-\frac{9326296}{81} \beta) L_{-3} L_{-2}^{6} L_{-1} + (-\frac{1718220067}{729}+\frac{565209697}{729} \beta) L_{-5} L_{-2}^{5} L_{-1}  \\
\lb + (\frac{611230072}{243}+\frac{3131873176}{1215} \beta) L_{-4} L_{-3} L_{-2}^{4} L_{-1} + (\frac{16790069291}{648}-\frac{311417113}{72} \beta) L_{-7} L_{-2}^{4} L_{-1}  \\
\lb + (\frac{508133378}{243}-\frac{522868522}{1215} \beta) L_{-3}^{3} L_{-2}^{3} L_{-1} + (-\frac{3350726791}{243}+\frac{16970704619}{1215} \beta) L_{-6} L_{-3} L_{-2}^{3} L_{-1}  \\
\lb + (-\frac{6183346072}{243}-\frac{12019334032}{1215} \beta) L_{-5} L_{-4} L_{-2}^{3} L_{-1} + (\frac{537159191}{648}-\frac{2545333679}{3240} \beta) L_{-9} L_{-2}^{3} L_{-1}  \\
\lb + (-\frac{7490314333}{243}-\frac{5707344769}{1215} \beta) L_{-5} L_{-3}^{2} L_{-2}^{2} L_{-1} + (\frac{4251155539}{81}-\frac{2417418029}{405} \beta) L_{-4}^{2} L_{-3} L_{-2}^{2} L_{-1}  \\
\lb + (-\frac{13774642501}{486}+\frac{47574031139}{2430} \beta) L_{-8} L_{-3} L_{-2}^{2} L_{-1} + (-\frac{30597896944}{81}+\frac{7602405812}{405} \beta) L_{-7} L_{-4} L_{-2}^{2} L_{-1}  \\
\lb + (-\frac{1818452026}{81}+\frac{2487870947}{405} \beta) L_{-7}^{2} L_{-1}^{2} + (-\frac{598093659731}{1944}-\frac{118832263337}{9720} \beta) L_{-11} L_{-2}^{2} L_{-1}  \\
\lb + (-\frac{1127428304}{81}+\frac{2005731856}{405} \beta) L_{-4} L_{-3}^{3} L_{-2} L_{-1} + (\frac{8937898306}{81}-\frac{9374211074}{405} \beta) L_{-7} L_{-3}^{2} L_{-2} L_{-1}  \\
\lb + (-\frac{2175490966}{81}-\frac{7301230378}{405} \beta) L_{-6} L_{-4} L_{-3} L_{-2} L_{-1} + (\frac{5686048555}{486}+\frac{71774155933}{2430} \beta) L_{-5}^{2} L_{-3} L_{-2} L_{-1}  \\
\lb + (\frac{5640483061}{162}-\frac{25935548597}{810} \beta) L_{-10} L_{-3} L_{-2} L_{-1} + (\frac{2068510520}{81}-\frac{175247464}{405} \beta) L_{-5} L_{-4}^{2} L_{-2} L_{-1}  \\
\lb + (\frac{30251263}{18}-\frac{780131773}{270} \beta) L_{-9} L_{-4} L_{-2} L_{-1} + (\frac{1652402846}{81}-\frac{13911205078}{405} \beta) L_{-8} L_{-5} L_{-2} L_{-1}  \\
\lb + (-\frac{2116132367}{27}+\frac{3248810563}{135} \beta) L_{-7} L_{-6} L_{-2} L_{-1} + (-\frac{8015375435}{648}-\frac{40699798345}{648} \beta) L_{-13} L_{-2} L_{-1}  \\
\lb + (-\frac{124276184}{81}-\frac{369101384}{405} \beta) L_{-3}^{5} L_{-1} + (\frac{474109532}{27}+\frac{439356476}{45} \beta) L_{-6} L_{-3}^{3} L_{-1}  \\
\lb + (\frac{813558668}{81}-\frac{5307013348}{405} \beta) L_{-5} L_{-4} L_{-3}^{2} L_{-1} + (-\frac{1380887179}{54}+\frac{6613469743}{270} \beta) L_{-9} L_{-3}^{2} L_{-1}  \\
\lb + (-\frac{185916220}{9}+\frac{20241916}{3} \beta) L_{-4}^{3} L_{-3} L_{-1} + (\frac{242801510}{9}-\frac{4142537918}{135} \beta) L_{-8} L_{-4} L_{-3} L_{-1}  \\
\lb + (-\frac{2519102860}{81}+\frac{1743712978}{81} \beta) L_{-7} L_{-5} L_{-3} L_{-1} + (-\frac{48696580}{81}-\frac{10819957036}{405} \beta) L_{-6}^{2} L_{-3} L_{-1}  \\
\lb + (\frac{3430860452}{81}+\frac{18602113922}{405} \beta) L_{-12} L_{-3} L_{-1} + (\frac{7433166527}{54}-\frac{2647655137}{270} \beta) L_{-7} L_{-4}^{2} L_{-1}  \\
\lb + (-\frac{4287087352}{81}+\frac{2955645208}{81} \beta) L_{-6} L_{-5} L_{-4} L_{-1} + (-\frac{887650963}{54}-\frac{8540950477}{270} \beta) L_{-11} L_{-4} L_{-1}  \\
\lb + (\frac{1877772049}{243}-\frac{1317572437}{243} \beta) L_{-5}^{3} L_{-1} + (\frac{10351048265}{81}+\frac{14282331911}{405} \beta) L_{-10} L_{-5} L_{-1}  \\
\lb + (\frac{382719419}{9}-\frac{178157449}{3} \beta) L_{-9} L_{-6} L_{-1} + (-\frac{45624853453}{324}+\frac{38349308507}{1620} \beta) L_{-8} L_{-7} L_{-1}  \\
\lb + (-\frac{45585350849}{162}-\frac{100134503837}{3240} \beta) L_{-15} L_{-1} + (\frac{22715000}{243}+\frac{4543000}{243} \beta) L_{-2}^{8}  \\
\lb + (-\frac{2052937000}{729}-\frac{548807000}{729} \beta) L_{-4} L_{-2}^{6} + (\frac{1729008875}{1458}+\frac{700883575}{1458} \beta) L_{-3}^{2} L_{-2}^{5}  \\
\lb + (\frac{31484615917}{2916}+\frac{13690595345}{2916} \beta) L_{-6} L_{-2}^{5} + (-\frac{141033886615}{5832}-\frac{54089482475}{5832} \beta) L_{-5} L_{-3} L_{-2}^{4}  \\
\lb + (\frac{16209468985}{648}+\frac{4917725989}{648} \beta) L_{-4}^{2} L_{-2}^{4} + (-\frac{69278656301}{1296}-\frac{5818122955}{432} \beta) L_{-8} L_{-2}^{4} \\
\lb  + (\frac{327789298}{243}-\frac{182443906}{243} \beta) L_{-4} L_{-3}^{2} L_{-2}^{3} + (\frac{579707603}{54}+\frac{1499507321}{162} \beta) L_{-7} L_{-3} L_{-2}^{3}  \\
\lb + (-\frac{71894257961}{486}-\frac{27674494435}{486} \beta) L_{-6} L_{-4} L_{-2}^{3} + (\frac{529286724439}{5832}+\frac{171551292269}{5832} \beta) L_{-5}^{2} L_{-2}^{3} \\
\lb  + (-\frac{219343338799}{1944}-\frac{48802331837}{1944} \beta) L_{-10} L_{-2}^{3} + (-\frac{752201776}{243}-\frac{99365984}{243} \beta) L_{-3}^{4} L_{-2}^{2}  \\
\lb + (\frac{3167581984}{81}+\frac{164797772}{81} \beta) L_{-6} L_{-3}^{2} L_{-2}^{2} + (\frac{7333242724}{243}+\frac{5959030124}{243} \beta) L_{-5} L_{-4} L_{-3} L_{-2}^{2}  \\
\lb + (-\frac{3744239456}{81}-\frac{4563477511}{162} \beta) L_{-9} L_{-3} L_{-2}^{2} + (-\frac{4568992472}{81}-\frac{1329441208}{81} \beta) L_{-4}^{3} L_{-2}^{2}  \\
\lb + (\frac{98011788952}{243}+\frac{23350859840}{243} \beta) L_{-8} L_{-4} L_{-2}^{2} + (-\frac{20981999923}{243}-\frac{24546873745}{486} \beta) L_{-7} L_{-5} L_{-2}^{2}  \\
\lb + (-\frac{12537042125}{243}+\frac{2293822643}{243} \beta) L_{-6}^{2} L_{-2}^{2} + (\frac{148867854119}{972}+\frac{25735300993}{972} \beta) L_{-12} L_{-2}^{2}  \\
\lb + (\frac{3397474222}{243}+\frac{592140590}{243} \beta) L_{-5} L_{-3}^{3} L_{-2} + (-\frac{153629552}{81}-\frac{157430272}{81} \beta) L_{-4}^{2} L_{-3}^{2} L_{-2}  \\
\lb + (-\frac{18698585200}{243}-\frac{3030629912}{243} \beta) L_{-8} L_{-3}^{2} L_{-2} + (\frac{332973250}{81}+\frac{334271030}{81} \beta) L_{-7} L_{-4} L_{-3} L_{-2}  \\
\lb + (\frac{4006295476}{243}+\frac{1534401548}{243} \beta) L_{-6} L_{-5} L_{-3} L_{-2} + (-\frac{39235245287}{243}-\frac{11785966714}{243} \beta) L_{-11} L_{-3} L_{-2}  \\
\lb + (\frac{3943923671}{27}+51922221 \beta) L_{-6} L_{-4}^{2} L_{-2} + (-\frac{56023386371}{486}-\frac{17628497017}{486} \beta) L_{-5}^{2} L_{-4} L_{-2}  \\
\lb + (\frac{18665820475}{162}+\frac{4831631945}{162} \beta) L_{-10} L_{-4} L_{-2} + (\frac{132951996853}{486}+\frac{49107342479}{486} \beta) L_{-9} L_{-5} L_{-2}  \\
\lb + (-\frac{2535966295}{18}-\frac{424001521}{6} \beta) L_{-8} L_{-6} L_{-2} + (-\frac{3527974355}{324}-\frac{2781374389}{324} \beta) L_{-7}^{2} L_{-2}  \\
\lb + (-\frac{256267624133}{486}-\frac{53727010357}{486} \beta) L_{-14} L_{-2} + (\frac{62662600}{81}+\frac{33837512}{81} \beta) L_{-4} L_{-3}^{4} \\
\lb  + (\frac{12081754}{81}-\frac{386061646}{81} \beta) L_{-7} L_{-3}^{3} + (-\frac{1235827894}{81}+\frac{324696958}{81} \beta) L_{-6} L_{-4} L_{-3}^{2}  \\
\lb + (-\frac{734163464}{81}-\frac{27208715}{81} \beta) L_{-5}^{2} L_{-3}^{2} + (\frac{1700122117}{54}-\frac{141939863}{18} \beta) L_{-10} L_{-3}^{2}  \\
\lb + (-\frac{644584063}{162}-\frac{1563558899}{162} \beta) L_{-5} L_{-4}^{2} L_{-3} + (-\frac{2365946831}{81}+\frac{391994255}{81} \beta) L_{-9} L_{-4} L_{-3}  \\
\lb + (\frac{41998314095}{324}+\frac{14424238267}{324} \beta) L_{-8} L_{-5} L_{-3} + (-\frac{5696336410}{81}-\frac{1244830214}{81} \beta) L_{-7} L_{-6} L_{-3}  \\
\lb + (-\frac{50875270591}{648}-\frac{5501974607}{162} \beta) L_{-13} L_{-3} + (\frac{141665042}{9}+\frac{45794666}{9} \beta) L_{-4}^{4}  \\
\lb + (-\frac{16720323782}{81}  -\frac{4469878486}{81} \beta) L_{-8} L_{-4}^{2} + (\frac{4719078932}{81}+\frac{1981388674}{81} \beta) L_{-7} L_{-5} L_{-4}  \\
\lb + (\frac{3329574508}{27}+\frac{774237704}{27} \beta) L_{-6}^{2} L_{-4} + (-\frac{24444294973}{81}-\frac{7245075107}{81} \beta) L_{-12} L_{-4}  \\
\lb + (-\frac{17795208008}{243}-\frac{5697658762}{243} \beta) L_{-6} L_{-5}^{2} + (\frac{172133662723}{648}+\frac{2558624480}{27} \beta) L_{-11} L_{-5}  \\
\lb + (-\frac{2482312006}{81}-\frac{359398637}{81} \beta) L_{-10} L_{-6} + (-\frac{11445305005}{324}-\frac{11100560395}{648} \beta) L_{-9} L_{-7}  \\
\lb + (\frac{88700463031}{486}+\frac{21332312435}{486} \beta) L_{-8}^{2} + (-\frac{273742696867}{324}-\frac{76299006809}{324} \beta) L_{-16} \Big) \; \cet{h-l} \\
\lb +  \Big( L_{-1}^{14}  -\frac{511}{6} L_{-2} L_{-1}^{12} + 273 L_{-3} L_{-1}^{11} + \frac{31031}{12} L_{-2}^{2} L_{-1}^{10}  -3575 L_{-4} L_{-1}^{10}  -\frac{71357}{6} L_{-3} L_{-2} L_{-1}^{9}  \\
\lb + \frac{156871}{9} L_{-5} L_{-1}^{9}  -\frac{7702123}{216} L_{-2}^{3} L_{-1}^{8} + \frac{2429141}{18} L_{-4} L_{-2} L_{-1}^{8}  -\frac{29315}{18} L_{-3}^{2} L_{-1}^{8}  \\
\lb  -\frac{330200}{9} L_{-6} L_{-1}^{8} + \frac{3232229}{18} L_{-3} L_{-2}^{2} L_{-1}^{7}  -\frac{14411306}{27} L_{-5} L_{-2} L_{-1}^{7}  -\frac{190736}{3} L_{-4} L_{-3} L_{-1}^{7}  \\
\lb + \frac{1113553}{12} L_{-7} L_{-1}^{7} + \frac{299830531}{1296} L_{-2}^{4} L_{-1}^{6}  -\frac{87916465}{54} L_{-4} L_{-2}^{2} L_{-1}^{6}  -\frac{1101737}{54} L_{-3}^{2} L_{-2} L_{-1}^{6}  \\
\lb + \frac{25678282}{27} L_{-6} L_{-2} L_{-1}^{6} + \frac{11453869}{108} L_{-5} L_{-3} L_{-1}^{6} + \frac{1494307}{3} L_{-4}^{2} L_{-1}^{6}  -\frac{18846499}{18} L_{-8} L_{-1}^{6}  \\
\lb  -\frac{120020173}{108} L_{-3} L_{-2}^{3} L_{-1}^{5} + \frac{9826873}{2} L_{-5} L_{-2}^{2} L_{-1}^{5} + \frac{12821156}{9} L_{-4} L_{-3} L_{-2} L_{-1}^{5}  \\
\lb  -\frac{52971385}{24} L_{-7} L_{-2} L_{-1}^{5} + \frac{858578}{9} L_{-3}^{3} L_{-1}^{5} + \frac{340520}{9} L_{-6} L_{-3} L_{-1}^{5}  -\frac{16430486}{9} L_{-5} L_{-4} L_{-1}^{5}  \\
\lb + \frac{53355355}{18} L_{-9} L_{-1}^{5}  -\frac{563493749}{864} L_{-2}^{5} L_{-1}^{4} + \frac{779378095}{108} L_{-4} L_{-2}^{3} L_{-1}^{4} + \frac{1494401}{3} L_{-3}^{2} L_{-2}^{2} L_{-1}^{4}  \\
\lb  -\frac{20491817}{3} L_{-6} L_{-2}^{2} L_{-1}^{4}  -\frac{165034463}{72} L_{-5} L_{-3} L_{-2} L_{-1}^{4}  -\frac{118077817}{18} L_{-4}^{2} L_{-2} L_{-1}^{4}  \\
\lb + \frac{558714659}{36} L_{-8} L_{-2} L_{-1}^{4}  -\frac{5919766}{9} L_{-4} L_{-3}^{2} L_{-1}^{4} + \frac{173851973}{72} L_{-7} L_{-3} L_{-1}^{4}  \\
\lb  -\frac{3001604}{3} L_{-6} L_{-4} L_{-1}^{4} + \frac{209551}{3} L_{-5}^{2} L_{-1}^{4} + \frac{19970203}{9} L_{-10} L_{-1}^{4} + \frac{370750733}{144} L_{-3} L_{-2}^{4} L_{-1}^{3}  \\
\lb  -\frac{88749761}{6} L_{-5} L_{-2}^{3} L_{-1}^{3}  -\frac{69331496}{9} L_{-4} L_{-3} L_{-2}^{2} L_{-1}^{3} + \frac{588707659}{48} L_{-7} L_{-2}^{2} L_{-1}^{3}  \\
\lb  -\frac{47120023}{54} L_{-3}^{3} L_{-2} L_{-1}^{3} + \frac{4427710}{3} L_{-6} L_{-3} L_{-2} L_{-1}^{3} + \frac{149796614}{9} L_{-5} L_{-4} L_{-2} L_{-1}^{3}  \\
\lb  -\frac{1533491879}{54} L_{-9} L_{-2} L_{-1}^{3} + \frac{52642805}{36} L_{-5} L_{-3}^{2} L_{-1}^{3} + 2193925 L_{-4}^{2} L_{-3} L_{-1}^{3} \\
\lb  -\frac{244608971}{18} L_{-8} L_{-3} L_{-1}^{3}  -\frac{68361415}{12} L_{-7} L_{-4} L_{-1}^{3} + \frac{104958070}{9} L_{-6} L_{-5} L_{-1}^{3}   \\
\lb-\frac{3755911505}{144} L_{-11} L_{-1}^{3} + \frac{359158375}{576} L_{-2}^{6} L_{-1}^{2}  -\frac{1422954571}{144} L_{-4} L_{-2}^{4} L_{-1}^{2}  -\frac{110349109}{72} L_{-3}^{2} L_{-2}^{3} L_{-1}^{2} \\
\lb + \frac{242776861}{18} L_{-6} L_{-2}^{3} L_{-1}^{2} + \frac{417557591}{48} L_{-5} L_{-3} L_{-2}^{2} L_{-1}^{2} + \frac{215671777}{12} L_{-4}^{2} L_{-2}^{2} L_{-1}^{2} \\
\lb  -\frac{1133758667}{24} L_{-8} L_{-2}^{2} L_{-1}^{2} + \frac{9402171}{2} L_{-4} L_{-3}^{2} L_{-2} L_{-1}^{2}  -\frac{46173362}{3} L_{-7} L_{-3} L_{-2} L_{-1}^{2} \\
\lb + \frac{8619874}{3} L_{-6} L_{-4} L_{-2} L_{-1}^{2}  -346185 L_{-5}^{2} L_{-2} L_{-1}^{2}  -12764834 L_{-10} L_{-2} L_{-1}^{2} + \frac{781726}{9} L_{-3}^{4} L_{-1}^{2} \\
\lb  -\frac{6265157}{3} L_{-6} L_{-3}^{2} L_{-1}^{2}  -\frac{17503249}{4} L_{-5} L_{-4} L_{-3} L_{-1}^{2} + \frac{851252395}{72} L_{-9} L_{-3} L_{-1}^{2} \\
\lb  -\frac{10377505}{3} L_{-4}^{3} L_{-1}^{2} + \frac{88750081}{2} L_{-8} L_{-4} L_{-1}^{2}  -\frac{107002967}{48} L_{-7} L_{-5} L_{-1}^{2}  -15108105 L_{-6}^{2} L_{-1}^{2} \\
\lb + \frac{421427881}{12} L_{-12} L_{-1}^{2}  -\frac{429866875}{288} L_{-3} L_{-2}^{5} L_{-1} + \frac{1485768463}{144} L_{-5} L_{-2}^{4} L_{-1} \\
\lb + \frac{75890573}{9} L_{-4} L_{-3} L_{-2}^{3} L_{-1}  -\frac{3917780737}{288} L_{-7} L_{-2}^{3} L_{-1} + \frac{47264245}{36} L_{-3}^{3} L_{-2}^{2} L_{-1} \\
\lb  -\frac{14009867}{3} L_{-6} L_{-3} L_{-2}^{2} L_{-1}  -\frac{47369767}{2} L_{-5} L_{-4} L_{-2}^{2} L_{-1} + \frac{3093452827}{72} L_{-9} L_{-2}^{2} L_{-1} \\
\lb  -\frac{114347681}{24} L_{-5} L_{-3}^{2} L_{-2} L_{-1}  -\frac{46986829}{6} L_{-4}^{2} L_{-3} L_{-2} L_{-1} + \frac{171211201}{4} L_{-8} L_{-3} L_{-2} L_{-1} \\
\lb + \frac{452772625}{24} L_{-7} L_{-4} L_{-2} L_{-1}  -\frac{96873449}{3} L_{-6} L_{-5} L_{-2} L_{-1} + \frac{2348340155}{32} L_{-11} L_{-2} L_{-1} \\
\lb  -\frac{8346652}{9} L_{-4} L_{-3}^{3} L_{-1} + \frac{36927503}{12} L_{-7} L_{-3}^{2} L_{-1} + 1760132 L_{-6} L_{-4} L_{-3} L_{-1}-\frac{41087977}{12} L_{-6} L_{-2}^{4}  \\
\lb + \frac{2278813}{12} L_{-5}^{2} L_{-3} L_{-1} + 12560898 L_{-10} L_{-3} L_{-1} + 5520381 L_{-5} L_{-4}^{2} L_{-1} + \frac{16051875}{32} L_{-3}^{2} L_{-2}^{4}  \\
\lb  -\frac{504079789}{18} L_{-9} L_{-4} L_{-1}  -\frac{114043262}{3} L_{-8} L_{-5} L_{-1} + 17921683 L_{-7} L_{-6} L_{-1} + \frac{178458875}{96} L_{-4} L_{-2}^{5} \\
\lb + \frac{3300324455}{144} L_{-13} L_{-1} -\frac{11116875}{128} L_{-2}^{7} -\frac{111748133}{32} L_{-5} L_{-3} L_{-2}^{3}  -\frac{136597993}{24} L_{-4}^{2} L_{-2}^{3} \\
\lb + \frac{262660881}{16} L_{-8} L_{-2}^{3}  -\frac{11084647}{4} L_{-4} L_{-3}^{2} L_{-2}^{2} + \frac{273622811}{32} L_{-7} L_{-3} L_{-2}^{2}  -22330 L_{-6} L_{-4} L_{-2}^{2} \\
\lb + \frac{690417}{4} L_{-5}^{2} L_{-2}^{2} + \frac{69379331}{12} L_{-10} L_{-2}^{2}  -\frac{1549555}{12} L_{-3}^{4} L_{-2} + \frac{4463795}{2} L_{-6} L_{-3}^{2} L_{-2} \\
\lb + \frac{43957095}{8} L_{-5} L_{-4} L_{-3} L_{-2}  -\frac{249418197}{16} L_{-9} L_{-3} L_{-2} + \frac{60703909}{18} L_{-4}^{3} L_{-2} -\frac{17478125}{4} L_{-8} L_{-3}^{2} \\
\lb  -\frac{531624373}{12} L_{-8} L_{-4} L_{-2} + \frac{245476273}{96} L_{-7} L_{-5} L_{-2} + \frac{90955697}{6} L_{-6}^{2} L_{-2}  -\frac{844913671}{24} L_{-12} L_{-2} \\
\lb + \frac{8812909}{24} L_{-5} L_{-3}^{3} + \frac{6587483}{6} L_{-4}^{2} L_{-3}^{2}   -\frac{103274297}{24} L_{-7} L_{-4} L_{-3} + 2522646 L_{-6} L_{-5} L_{-3}  \\
\lb -\frac{210841967}{24} L_{-11} L_{-3} + \frac{3034556}{3} L_{-6} L_{-4}^{2}  -525399 L_{-5}^{2} L_{-4}  -17243744 L_{-10} L_{-4}  \\
\lb + \frac{64762145}{6} L_{-9} L_{-5} + 6664196 L_{-8} L_{-6}  -\frac{400057469}{96} L_{-7}^{2}  -\frac{254131871}{12} L_{-14} \Big) \; \cet{h;1} \, .
\eeq

%%%%%%%%%

\section{Explicit nullvectors in the bulk of $\mathbf{c_{2,3} = 0}$\label{app2}}

In the following we give the explicit form of the nullvector of type {\bf E} 
which has a Jordan cell at $h=1$, lowest weight $h=0$ and appears at
level $12$. Again, for the sake of brevity, we have 
set the overall normalisation to $1$
and also eliminated any further freedom by setting any further free parameter to $0$;
again, the parameter $\beta$ remains as it is a parameter of the representation as introduced in
section \ref{section3}:
\beq
\lb \Big( -\frac{44800}{27} L_{-3} L_{-2}^{4} L_{-1} + (\frac{358528}{27}-\frac{4000}{9} \beta) L_{-5} L_{-2}^{3} L_{-1} + (\frac{117920}{27}+\frac{49600}{9} \beta) L_{-4} L_{-3} L_{-2}^{2} L_{-1} \\
\lb  + (-\frac{1543136}{81}-\frac{814208}{81} \beta) L_{-7} L_{-2}^{2} L_{-1} + (-\frac{572912}{81}-\frac{100000}{27} \beta) L_{-3}^{3} L_{-2} L_{-1} \\
\lb + (\frac{620576}{9}-\frac{217408}{9} \beta) L_{-6} L_{-3} L_{-2} L_{-1} + (-\frac{1214624}{27}+\frac{243424}{9} \beta) L_{-5} L_{-4} L_{-2} L_{-1} \\
\lb + (\frac{3934496}{81}-\frac{341888}{27} \beta) L_{-9} L_{-2} L_{-1} + (\frac{551312}{27}+\frac{396352}{9} \beta) L_{-5} L_{-3}^{2} L_{-1} \\
\lb + (-\frac{91424}{9}-\frac{116416}{3} \beta) L_{-4}^{2} L_{-3} L_{-1} + (-\frac{619904}{27}-1408 \beta) L_{-8} L_{-3} L_{-1} \\
\lb + (\frac{548672}{9}+\frac{769984}{9} \beta) L_{-7} L_{-4} L_{-1} + (-\frac{4270240}{81}-\frac{532160}{9} \beta) L_{-6} L_{-5} L_{-1} \\
\lb + (\frac{3200}{9}-\frac{44800}{27} \beta) L_{-4} L_{-2}^{4} + (\frac{10400}{3}+\frac{11200}{9} \beta) L_{-3}^{2} L_{-2}^{3} + (-\frac{2984320}{243}+\frac{585856}{27} \beta) L_{-6} L_{-2}^{3} \\
\lb + (-\frac{1611424}{81}-\frac{1122080}{27} \beta) L_{-5} L_{-3} L_{-2}^{2}  + (-\frac{63488}{27}+36864 \beta) L_{-4} L_{-3}^{2} L_{-2} \\
\lb + (\frac{309376}{9}+\frac{72832}{3} \beta) L_{-11} L_{-1} + (\frac{571936}{81}+\frac{304064}{9} \beta) L_{-8} L_{-2}^{2}+ (\frac{195488}{27}-\frac{9920}{3} \beta) L_{-4}^{2} L_{-2}^{2} \\
\lb + (\frac{229504}{81}-\frac{5155328}{81} \beta) L_{-7} L_{-3} L_{-2} + (-\frac{1020352}{81}-\frac{1000192}{9} \beta) L_{-6} L_{-4} L_{-2} \\
\lb + (\frac{451616}{27}+\frac{709312}{9} \beta) L_{-5}^{2} L_{-2} + (-\frac{200192}{9}-\frac{279040}{3} \beta) L_{-10} L_{-2} + (\frac{40096}{81}-\frac{318016}{27} \beta) L_{-3}^{4} \\
\lb + (-\frac{693152}{81}+\frac{1014976}{9} \beta) L_{-6} L_{-3}^{2}  + (-\frac{1646560}{81}+\frac{3016960}{27} \beta) L_{-9} L_{-3} + (\frac{1521088}{81}-\frac{762496}{9} \beta) L_{-6}^{2} \\
\lb + (-\frac{35168}{9}+\frac{95168}{3} \beta) L_{-4}^{3} + (\frac{128672}{27}-\frac{892096}{9} \beta) L_{-8} L_{-4} + (-\frac{758240}{81} +\frac{6217312}{81} \beta) L_{-7} L_{-5} \\
\lb + (\frac{306176}{27}-\frac{234080}{3} \beta) L_{-5} L_{-4} L_{-3}  + (-\frac{947200}{27}+\frac{758080}{9} \beta) L_{-12} \Big)  \; \cet{h-l} \\
\lb  +  \Big( L_{-1}^{11}  -44 L_{-2} L_{-1}^{9} + 88 L_{-3} L_{-1}^{8} + \frac{1804}{3} L_{-2}^{2} L_{-1}^{7}  -836 L_{-4} L_{-1}^{7}  -1320 L_{-3} L_{-2} L_{-1}^{6} \\
\lb  + \frac{16456}{9} L_{-5} L_{-1}^{6}  -\frac{85184}{27} L_{-2}^{3} L_{-1}^{5} + \frac{107536}{9} L_{-4} L_{-2} L_{-1}^{5}  -\frac{12232}{9} L_{-3}^{2} L_{-1}^{5} + \frac{4640}{9} L_{-6} L_{-1}^{5} \\
\lb  + \frac{52448}{9} L_{-3} L_{-2}^{2} L_{-1}^{4}  -\frac{168112}{9} L_{-5} L_{-2} L_{-1}^{4} + 720 L_{-4} L_{-3} L_{-1}^{4}  -1360 L_{-7} L_{-1}^{4} + \frac{53504}{9} L_{-2}^{4} L_{-1}^{3}  \\
\lb  -\frac{373888}{9} L_{-4} L_{-2}^{2} L_{-1}^{3} + \frac{59072}{9} L_{-3}^{2} L_{-2} L_{-1}^{3}  -\frac{13760}{9} L_{-6} L_{-2} L_{-1}^{3}  -\frac{56896}{9} L_{-5} L_{-3} L_{-1}^{3}  \\
\lb + 15520 L_{-4}^{2} L_{-1}^{3}  -\frac{214816}{9} L_{-8} L_{-1}^{3}  -\frac{22784}{3} L_{-3} L_{-2}^{3} L_{-1}^{2} + \frac{120320}{3} L_{-5} L_{-2}^{2} L_{-1}^{2}  \\
\lb  -\frac{7040}{3} L_{-4} L_{-3} L_{-2} L_{-1}^{2} + \frac{3904}{3} L_{-7} L_{-2} L_{-1}^{2} + \frac{22016}{9} L_{-3}^{3} L_{-1}^{2} + 3008 L_{-6} L_{-3} L_{-1}^{2}   \\
\lb -\frac{46016}{3} L_{-5} L_{-4} L_{-1}^{2} + \frac{161152}{9} L_{-9} L_{-1}^{2}  -\frac{8192}{3} L_{-2}^{5} L_{-1} + 29696 L_{-4} L_{-2}^{3} L_{-1} \\
\lb   -4672 L_{-3}^{2} L_{-2}^{2} L_{-1}  -\frac{2048}{3} L_{-6} L_{-2}^{2} L_{-1} + \frac{27392}{3} L_{-5} L_{-3} L_{-2} L_{-1}  -\frac{94208}{3} L_{-4}^{2} L_{-2} L_{-1} \\
\lb  + 53120 L_{-8} L_{-2} L_{-1} + \frac{1984}{3} L_{-4} L_{-3}^{2} L_{-1} + \frac{24832}{3} L_{-7} L_{-3} L_{-1}  -\frac{45952}{3} L_{-6} L_{-4} L_{-1}  \\
\lb  -\frac{3776}{3} L_{-5}^{2} L_{-1} + 22784 L_{-10} L_{-1} + \frac{4096}{3} L_{-3} L_{-2}^{4}  -9984 L_{-5} L_{-2}^{3} + 128 L_{-4} L_{-3} L_{-2}^{2} \\
\lb  + \frac{3328}{3} L_{-7} L_{-2}^{2}  -\frac{3968}{3} L_{-3}^{3} L_{-2}  -\frac{5632}{3} L_{-6} L_{-3} L_{-2} + \frac{33536}{3} L_{-5} L_{-4} L_{-2}  \\
\lb  -\frac{39808}{3} L_{-9} L_{-2} + \frac{4480}{3} L_{-5} L_{-3}^{2}  -\frac{3200}{3} L_{-4}^{2} L_{-3}  -4480 L_{-8} L_{-3} \\
\lb   -\frac{6656}{3} L_{-7} L_{-4} + 8064 L_{-6} L_{-5}  -\frac{48256}{3} L_{-11} \Big) \; \cet{h;1} \; .
\eeq

The explicit form of the nullvector of type {\bf F}
which has a Jordan cell at $h=2$, lowest weight $h=0$
and also appears at level $12$ is given below.
As we need the full beauty of this result in the
argument of section \ref{section3} any free parameter
appears as calculated (noted as $m_i$); only the overall normalisation
we have set to $1$:
\beq
\lb \Big( m_{76} L_{-1}^{12} + m_{75} L_{-2} L_{-1}^{10} + m_{74} L_{-3} L_{-1}^{9} + m_{73} L_{-2}^{2} L_{-1}^{8} + m_{72} L_{-4} L_{-1}^{8}  \\
\lb + m_{71} L_{-3} L_{-2} L_{-1}^{7} + m_{70} L_{-5} L_{-1}^{7} + m_{69} L_{-2}^{3} L_{-1}^{6} + m_{68} L_{-4} L_{-2} L_{-1}^{6} + m_{67} L_{-3}^{2} L_{-1}^{6}  \\
\lb + m_{66} L_{-6} L_{-1}^{6} + m_{65} L_{-3} L_{-2}^{2} L_{-1}^{5} + m_{64} L_{-5} L_{-2} L_{-1}^{5} + m_{63} L_{-4} L_{-3} L_{-1}^{5}  \\
\lb + m_{62} L_{-7} L_{-1}^{5} + m_{61} L_{-2}^{4} L_{-1}^{4} + m_{60} L_{-4} L_{-2}^{2} L_{-1}^{4} + m_{59} L_{-3}^{2} L_{-2} L_{-1}^{4} \\
\lb + m_{58} L_{-6} L_{-2} L_{-1}^{4} + m_{57} L_{-5} L_{-3} L_{-1}^{4} + m_{56} L_{-4}^{2} L_{-1}^{4} + m_{55} L_{-8} L_{-1}^{4}  \\
\lb + m_{54} L_{-3} L_{-2}^{3} L_{-1}^{3} + m_{53} L_{-5} L_{-2}^{2} L_{-1}^{3} + m_{52} L_{-4} L_{-3} L_{-2} L_{-1}^{3} + m_{51} L_{-7} L_{-2} L_{-1}^{3} \\
\lb + m_{50} L_{-3}^{3} L_{-1}^{3} + m_{49} L_{-6} L_{-3} L_{-1}^{3} + m_{48} L_{-5} L_{-4} L_{-1}^{3} + m_{47} L_{-9} L_{-1}^{3} + m_{46} L_{-2}^{5} L_{-1}^{2} \\
\lb + m_{45} L_{-4} L_{-2}^{3} L_{-1}^{2} + m_{44} L_{-3}^{2} L_{-2}^{2} L_{-1}^{2} + m_{43} L_{-6} L_{-2}^{2} L_{-1}^{2} + m_{42} L_{-5} L_{-3} L_{-2} L_{-1}^{2} \\
\lb + m_{41} L_{-4}^{2} L_{-2} L_{-1}^{2} + m_{40} L_{-8} L_{-2} L_{-1}^{2} + m_{39} L_{-4} L_{-3}^{2} L_{-1}^{2} + m_{38} L_{-7} L_{-3} L_{-1}^{2}  \\
\lb+ m_{37} L_{-6} L_{-4} L_{-1}^{2} + m_{36} L_{-5}^{2} L_{-1}^{2} + m_{35} L_{-10} L_{-1}^{2} + m_{34} L_{-3} L_{-2}^{4} L_{-1}  \\
\lb+ m_{33} L_{-5} L_{-2}^{3} L_{-1} + m_{32} L_{-4} L_{-3} L_{-2}^{2} L_{-1} + m_{31} L_{-7} L_{-2}^{2} L_{-1} + m_{30} L_{-3}^{3} L_{-2} L_{-1}  \\
\lb+ m_{29} L_{-6} L_{-3} L_{-2} L_{-1} + m_{28} L_{-5} L_{-4} L_{-2} L_{-1} + m_{27} L_{-9} L_{-2} L_{-1} + m_{26} L_{-5} L_{-3}^{2} L_{-1} \\
\lb + m_{25} L_{-4}^{2} L_{-3} L_{-1} + m_{24} L_{-8} L_{-3} L_{-1} + m_{23} L_{-7} L_{-4} L_{-1} + m_{22} L_{-6} L_{-5} L_{-1} + \\
\lb m_{21} L_{-11} L_{-1}  -4096 L_{-4} L_{-2}^{4} + 3072 L_{-3}^{2} L_{-2}^{3} + (12800-8192 \beta) L_{-6} L_{-2}^{3}  \\
\lb+ (-32960+14336 \beta) L_{-5} L_{-3} L_{-2}^{2} + (30976+6144 \beta) L_{-4}^{2} L_{-2}^{2} + (-25088-17920 \beta) L_{-8} L_{-2}^{2}  \\
\lb + (-4736-18432 \beta) L_{-4} L_{-3}^{2} L_{-2} + (18560+32640 \beta) L_{-7} L_{-3} L_{-2}  \\
\lb+ (-73216+50176 \beta) L_{-6} L_{-4} L_{-2} + (54464-36608 \beta) L_{-5}^{2} L_{-2}+ (-496+5760 \beta) L_{-3}^{4}  \\
\lb + (-18432+48128 \beta) L_{-10} L_{-2} + (12096-54784 \beta) L_{-6} L_{-3}^{2} + (2432+40448 \beta) L_{-5} L_{-4} L_{-3}  \\
\lb+ (-11648-59008 \beta) L_{-9} L_{-3} + (-\frac{22144}{3}-17408 \beta) L_{-4}^{3} + (27264+55040 \beta) L_{-8} L_{-4} \\
\lb + (-5120-39040 \beta) L_{-7} L_{-5} + (-4288+41472 \beta) L_{-6}^{2} + (31552-35328 \beta) L_{-12} \Big) \; \cet{h-l}  \\
\lb +  \Big( L_{-1}^{10}  -\frac{130}{3} L_{-2} L_{-1}^{8} + \frac{284}{3} L_{-3} L_{-1}^{7} + \frac{5152}{9} L_{-2}^{2} L_{-1}^{6}  -776 L_{-4} L_{-1}^{6}  -1488 L_{-3} L_{-2} L_{-1}^{5} \\
\lb + \frac{6232}{3} L_{-5} L_{-1}^{5}  -\frac{8320}{3} L_{-2}^{3} L_{-1}^{4} + \frac{29440}{3} L_{-4} L_{-2} L_{-1}^{4}  -\frac{2752}{3} L_{-3}^{2} L_{-1}^{4} + 640 L_{-6} L_{-1}^{4} \\
\lb + \frac{21376}{3} L_{-3} L_{-2}^{2} L_{-1}^{3}  -\frac{67264}{3} L_{-5} L_{-2} L_{-1}^{3} + \frac{800}{3} L_{-4} L_{-3} L_{-1}^{3}  -800 L_{-7} L_{-1}^{3} + 4096 L_{-2}^{4} L_{-1}^{2} \\
\lb  -23552 L_{-4} L_{-2}^{2} L_{-1}^{2} + 992 L_{-3}^{2} L_{-2} L_{-1}^{2}  -3840 L_{-6} L_{-2} L_{-1}^{2} + 8896 L_{-5} L_{-3} L_{-1}^{2} \\
\lb  -9024 L_{-8} L_{-1}^{2}  -10240 L_{-3} L_{-2}^{3} L_{-1} + 48768 L_{-5} L_{-2}^{2} L_{-1} + 4160 L_{-4} L_{-3} L_{-2} L_{-1} \\
\lb  -4992 L_{-7} L_{-2} L_{-1} + 18048 L_{-6} L_{-3} L_{-1}  -23552 L_{-5} L_{-4} L_{-1} + 30528 L_{-9} L_{-1} \\
\lb  -8192 L_{-4} L_{-2}^{3} + 10240 L_{-3}^{2} L_{-2}^{2} + 1024 L_{-6} L_{-2}^{2}  -60800 L_{-5} L_{-3} L_{-2} \\
\lb + 47104 L_{-4}^{2} L_{-2}  -30720 L_{-8} L_{-2}  -6144 L_{-4} L_{-3}^{2} + 15232 L_{-7} L_{-3}  \\
\lb -39424 L_{-6} L_{-4} + 36992 L_{-5}^{2}  -24576 L_{-10} \Big) \; \cet{h;1} \; .
\eeq

\end{appendix}

%%%%%%%%%%%%%%%%%%%%%%%%%%%%%%%%%%%%%%%%%%%%%

\bibliographystyle{JHEP}
\bibliography{holgerJHEP}

%%%%%%%%%%%%%%%%%%%%%%%%%%%%%%%%%%%%%%%%%%%%%
\end{document}